\documentclass[12pt]{article}
\usepackage{epsf}
\def\reference{\parskip 0pt\par\noindent\hangindent 0.5 truecm}
\def\kms{km ${\rm s}^{-1}$}

\def\wisk#1{\ifmmode{#1}\else{$#1$}\fi}

\def\kms{\ifmmode {\>{\rm km\ s}^{-1}}\else {km s$^{-1}$}\fi}
\def\gtrapprox{\;\lower 0.5ex\hbox{$\buildrel >
    \over \sim\ $}}             
\def\lessapprox{\;\lower 0.5ex\hbox{$\buildrel < \over \sim\ $}}
\def\etal{et al.\ }
\def\eg{e.g.,\ }
\def\ie{i.e.,\ }
\def\cf{cf.\ }
\def\htwo{\wisk{\rm H_2}}
\def\htwoo{\wisk{\rm H_2O}}
\def\ergpsec{\wisk{\rm erg\ s^{-1}}}
\def\pyr{\wisk{\rm yr^{-1}}}
\def\psec{\wisk{\rm sec^{-1}}}
\def\psqcm{\ifmmode {\>{\rm cm}^{-2}}\else {cm$^{-2}$}\fi}
\def\flux{\ifmmode {\>{\rm erg\ cm^{-2}\ s^{-1}}}\else {erg cm$^{-2}$
s$^{-1}$}\fi}
\def\pcubcm{\ifmmode {\>{\rm cm}^{-3}}\else {cm$^{-3}$}\fi}
\def\pcubpc{\ifmmode {\>{\rm pc}^{-3}}\else {pc$^{-3}$}\fi}
\def\xieff{\wisk{\xi_{\rm eff}}}
\def\phisat{\wisk{\Phi_{\rm sat}}}
\def\be{\begin{equation}}
\def\ee{\end{equation}}
\def\bea{\begin{eqnarray*}}
\def\eea{\end{eqnarray*}}
\def\rpcsq{\ifmmode {r_{\rm pc}^2}\else {$r_{\rm pc}^2$}\fi}
\def\rpc{\ifmmode {r_{\rm pc}}\else {$r_{\rm pc}$}\fi}
\def\fabs{\ifmmode {f_{\rm abs}}\else {$f_{\rm abs}$}\fi}
\def\msol{\ifmmode {\>M_\odot}\else {$M_\odot$}\fi}
\def\lsol{\ifmmode {\>L_\odot}\else {$L_\odot$}\fi}
\def\plotfiddle#1#2#3#4#5#6#7{\centering \leavevmode
\vbox to#2{\rule{0pt}{#2}}
\includegraphics{#1}}
\newcommand{\sm}{\scriptsize}
\newlength{\phantomdigit}
\newcommand{\z}{\hspace*{\phantomdigit}}

\begin{document}
%
%
\title{Powerful Water Masers in Active Galactic Nuclei}
%

\author{Philip R. Maloney
} 

\date{}
\maketitle

{\center
CASA, University of Colorado, Boulder, CO,
80309-0389\\maloney@casa.colorado.edu\\[3mm]}

\begin{abstract}
Luminous water maser emission in the $6_{16}-5_{23}$ line at 22~GHz
has been detected from two dozen galaxies. In all cases the emission
is confined to the nucleus and has been found only in AGN, in
particular, in Type 2 Seyferts and LINERs. I argue that most of the
observed megamaser sources are powered by X-ray irradiation of dense
gas by the central engine. After briefly reviewing the physics of
these X-Ray Dissociation Regions, I discuss in detail the observations
of the maser disk in NGC 4258, its implications, and compare
alternative models for the maser emission. I then discuss the
observations of the other sources that have been imaged with VLBI to
date, and how they do or do not fit into the framework of a thin,
rotating disk, as in NGC 4258. Finally, I briefly discuss future
prospects, especially the possibility of detecting other water maser
transitions.
\end{abstract}


\bigskip

%
%

\section{Extragalactic Water Masers}
Here I briefly review the discovery of powerful water maser sources in
other galaxies; more detailed reviews, by active participants in
the observation and interpretation of \htwoo\ megamasers using VLBI,
are provided in Moran \etal (1995) and Greenhill (2001).
\subsection{Water Maser Emission in Galactic Sources}
Maser emission in rotational lines of water -- usually the
$6_{16}\rightarrow 5_{23}$ transition at 22~GHz (1.35 cm) has been
observed from Galactic sources for more than three decades (Cheung
\etal 1969), from star-forming regions and late-type stars. The
population inversion necessary to produce maser emission in this
transition can be produced simply by collisions (\eg shocks), due to
the interactions between the rotational ``ladders'' of the \htwoo\
molecule (de Jong 1973; see Elitzur 1992 for a detailed recent
discussion). The observed luminosities are typically $L_{\rm H_2O}\sim
10^{-3}\lsol$, although the most luminous Galactic source, W49,
occasionally reaches $L_{\rm H_2O}\sim 1\lsol$ (Genzel \& Downes
1979)\footnote{Maser luminosities are traditionally given as the
isotropic value $4\pi D^2F$, where $D$ is the source distance and $F$
is the line flux. Since maser emission is highly anisotropic, this can
overestimate the true luminosity of an observed maser
feature. However, precisely because of the anisotropy, there are
almost certainly many maser features that are {\it not} beamed in our
direction, and so the isotropic luminosity based on the observed
features may provide a reasonable estimate of the actual maser
luminosity.}. Since the levels involved in the 22 GHz transition lie
at energies $E/k\approx 600$ K above the ground state, the gas
temperature must be at least a few hundred K in order to excite the
emission.

\subsection{Extragalactic Sources}
The first water masers to be detected in other galaxies (\eg M33:
Churchwell \etal 1977) resembled W49, with $L_{\rm H_2O}\sim 1\lsol$,
and were typically found in galactic disks, in regions of star
formation. These could be easily understood as extragalactic analogues
of the Galactic water maser sources. However, a new and unexpected
type of extragalactic \htwoo\ maser was discovered by dos Santos \&
Lepine (1979) in the edge-on spiral NGC 4945. With an isotropic
luminosity $L_{\rm H_2O}\sim 100\lsol$, this source -- which evidently
arose in the galactic nucleus -- was five orders of magnitude more
luminous than typical Galactic \htwoo\ masers, and was classed as a
water ``megamaser'' ($L\sim 10^6\times L_{\rm H_2O}$(Galactic)). Four
more megamasers were found over the next 5 years, including Circinus
(Gardner \& Whiteoak 1982), NGC 3079 (Henkel \etal 1984a,b) and NGC
1068 and NGC 4258 (Claussen, Heiligman, \& Lo 1984). Typically the
emission consisted of one to a few components with velocity widths
$\delta V \sim$ few \kms, spread over a total velocity range $\Delta
V\sim 100$ \kms\ about the systemic velocity of the galaxy $V_{\rm
sys}$; an example (NGC 4945) is shown in Figure 1. Additional surveys
over the next several years, mostly in late-type and infrared-luminous
galaxies, failed to detect any new sources (\eg Claussen \& Lo 1986;
Whiteoak \& Gardner 1986).

\begin{figure}
\plotfiddle{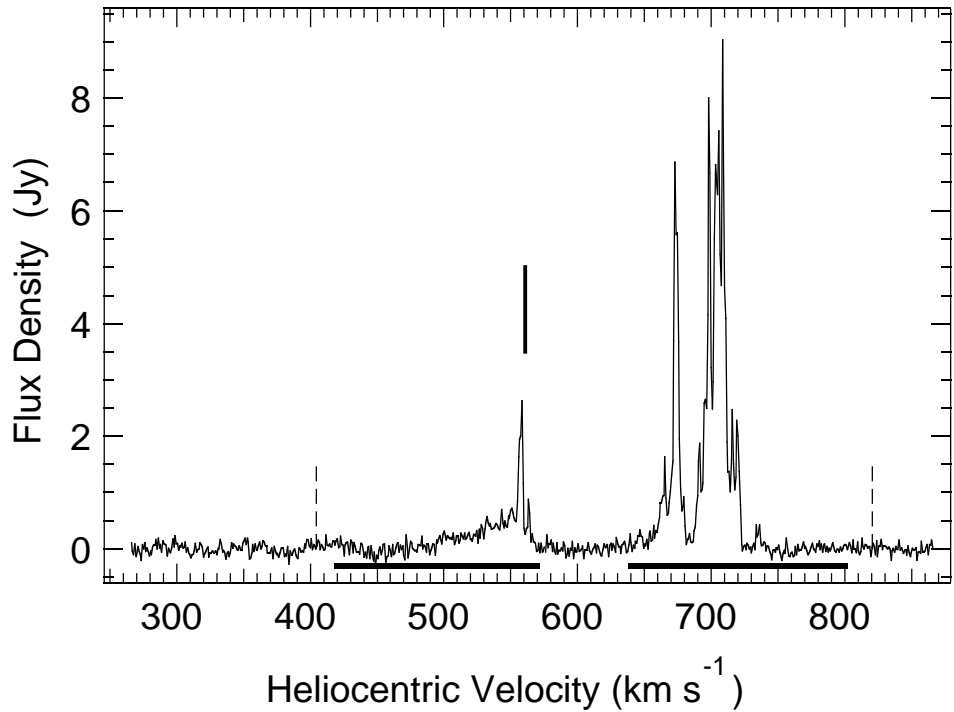}{3.4in}{0}{100}{100}{-320}{-290}
\caption{The spectrum of \htwoo\ maser emission from NGC 4945, showing
the spread of emission with respect to the systemic velocity
(indicated by the vertical bar). From Greenhill \etal 1997a.} 
\label{fig1}
\end{figure}

However, all 5 of the known megamasers shared common features, as
first noted by Claussen \etal (1984):
\begin{itemize}
\item They all were found in host galaxies with evidence of nuclear
activity: three were classified as Seyferts or LINERs, and three
showed unusual, extended radio emission or strong nuclear radio
continuum emission; 
\item The maser emission was centered on the nucleus;
\item In the cases of NGC 1068 and NGC 4258, the spatial diameter of
the emitting regions were constrained by interferometric
observations to be $d < 3.5$ and 
1.3~pc, respectively (Claussen \& Lo 1986).
\end{itemize}

Taking a cue from these features of the galaxies containing \htwoo\
megamasers, Braatz, Wilson, \& Henkel (1996, 1997), undertook a systematic
survey of AGN for H$_2$O megamasers. They observed 354 active galaxies
(both Seyferts and LINERs). Ten new megamasers were found, raising the
total to 16. An examination of the statistics in both distance-limited
($cz < 7000$ \kms) and sensitivity-limited ($L_{\rm H_2O} > 2\lsol$)
samples leads to a remarkable result:

\noindent In the distance-limited sample:
\begin{itemize}
\item{$7\%$ (10/141) Sy 2 detected}
\item{$7.5\%$ (5/67) LINERs detected}
\item{$0\%$ (0/57) Sy 1 detected}
\end{itemize}

\noindent For the sensitivity-limited sample, these percentages become:
\begin{itemize}
\item{$14\%$ (10/73) Sy 2 detected}
\item{$10\%$ (5/52) LINERs detected}
\item{$0\%$ (0/30) Sy 1 detected}
\end{itemize}
The fraction of H$_2$O megamasers in AGN is {\it not} small, provided
that Type 1 Seyferts are excluded. Hence either: (1) Sy 1 don't
possess nuclear molecular gas in which the physical conditions are
suitable for masing, or (2) the megamasers in 
Sy 1 galaxies are systematically beamed {\it away} from us.

Since the Braatz \etal survey, eight more megamasers have been
discovered\footnote{A database of galaxies that have been searched for
22 GHz water maser emission is maintained by Jim Braatz, and is
available at http://www.gb.nrao.edu/$\sim$jbraatz/H2O-list.html.},
bringing the current total to 24 (see the list in Moran, Greenhill, \&
Herrnstein 1999, plus Mrk 348 [Falcke \etal 2000b] and NGC 6240
[Hagiwara, Diamond, \& Miyoshi 2002; Nakai, Sato, \& Yamauchi
2002]). All of the sources are in galaxies classified as either Sy 2
(14/24) or LINERs (10/24). In all cases the emission is confined to
the nucleus (although in the case of the merger NGC 6240, the location
of the nucleus is somewhat problematic). The survey reported by
Greenhill \etal (2002) detected only one new source out of 131
galaxies searched; however, only about 50 of these sources are classed
as Sy 2 or LINERs. Nine megamaser sources have now been imaged using
VLBI; the results will be discussed in \S 4 and 5.

\section{The origin of water megamasers}
As mentioned above, the $6_{16}\rightarrow 5_{23}$ transition of
\htwoo\ at 22 GHz can be driven to mase by collisions, provided that
the densities and temperatures are high enough ($n_\htwo\gtrapprox
10^7 \pcubcm$, $T\gtrapprox 400$ K, respectively) to excite the levels
involved, and that the levels deviate from thermodynamic equilibrium,
which implies that $n_\htwo\lessapprox 10^{11} \pcubcm$, and that the
gas kinetic temperature is substantially higher than the temperature
of the radiation field in the mid- to far-infrared, where the pure
rotational transitions connecting the levels occur. In the Galactic
sources of water maser emission, namely, star-forming regions and the
atmospheres of late-type stars, there is little doubt that the
necessary conditions are produced by shock waves. However, the unique
association between water megamasers and active galactic nuclei
suggests that there is another mechanism at work in many if not all of
these sources. As I will show below, the best explanation for the
majority of the sources is likely to be maser emission arising in
dense gas that is being irradiated by X-rays from the AGN. Before
discussing the maser emission, I will first digress briefly to discuss
the characteristics of these {\it X-Ray Dissociation Regions}.

\subsection{X-Ray Dissociation Regions}
Consider a power-law spectrum incident on a layer of dense gas: after
spectral filtering through absorption in the gas, only X-rays, with
energies $E$ above a few hundred eV, are left. This is very
significant for active galactic nuclei, because for typical AGN
spectra, approximately equal amounts of energy per decade are
available. Hence the hard X-ray ($E\gtrapprox$ 1 keV) luminosity $L_x$
is roughly one-tenth of the bolometric luminosity $L_{bol}$. These
hard X-rays can have a profound influence on the physical state of the
ISM in a galaxy with an active nucleus, since photons with $E
\gtrapprox 1$ kev have small cross-sections for absorption, and hence
large mean free paths, but large $E$/photon, so that the local energy
input rates when the photons are absorbed are substantial. Much of the
locally deposited energy goes into heating the gas; the remainder goes
into ionization and excitation (Maloney, Hollenbach, \& Tielens
1996). For a typical ($f_\nu\propto\nu^{-\alpha}$, with $\alpha\sim
1$) AGN X-ray spectrum, the ionization and heating rates fall with the
column density $N$ only as $\sim N^{-1}$. This is in marked contrast
to the {\it photodissociation regions} (PDRs) produced by FUV photons
(Tielens \& Hollenbach 1985), in which the heating rates decline
exponentially with column density since the absorption is produced by
dust. This means that the column density in an XDR can be large, and
therefore that the volume around a luminous AGN occupied by the XDR
(\ie where X-rays dominate the ionization and heating rates) can be
substantial. For typical AGN spectra, the ionization and heating rates
are always dominated by the lowest energy photons that have yet to be
significantly attenuated. The physical state of the gas is described
to lowest order by an effective X-ray ionization parameter \xieff,
related to the usual ratio of X-ray flux to gas density, but including a
scaling with column density that accounts for the optical thickness of
the gas to the incident radiation. This scaling depends weakly on the
spectral index but is generally close to $N^{-1}$, as noted above (see
Maloney, Hollenbach, \& Tielens 1996 for details).

\subsection{Water Maser Emission from Dense X-Ray Irradiated Gas in AGN}
High-pressure XDRs prove to be ideal locations for the production of
luminous water maser emission (Neufeld, Maloney, \& Conger 1994,
hereafter NMC). Provided the gas temperature is above $\sim 400$ K but
is not so high ($T\sim 4000$ K) that H$_2$ is destroyed, the reaction
network 
\vskip 0.2in
\centerline{O + H$_2$ $\rightarrow$ OH + H}
\centerline{OH + H$_2$ $\rightarrow$ H$_2$O + H}
\vskip 0.2in
\noindent forms water very efficiently. This is also the approximate
temperature criterion for collisionally exciting the 22 GHz
transition, as discussed earlier. An X-ray-irradiated layer of dense
gas generally undergoes a phase transition from warm, atomic gas to
cooler, molecular gas. An example is shown in Figure 2. The
temperature plummets from $T\sim 6000$ K to $T\sim 1000$ K at the
transition, while the \htwo\ abundance climbs from $\sim$ a few
$\times 10^{-5}$ to $\sim$ ten percent. The water abundance relative to
hydrogen jumps from negligible values ($x_\htwoo\lessapprox 10^{-12}$)
to $\sim $ a few $\times 10^{-6}$, and then climbs to $\sim 10^{-4}$
before starting to decline as the temperature drops with increasing
column density. (The initial rise in temperature following the
transition is due to the effects of radiative trapping, which acts to
reduce the effective cooling rate per unit volume as the optical
depths in important cooling transitions increase; this effect is
eventually outweighed by the continued decline in the gas heating rate
with increasing $N$.)

\begin{figure}
\plotfiddle{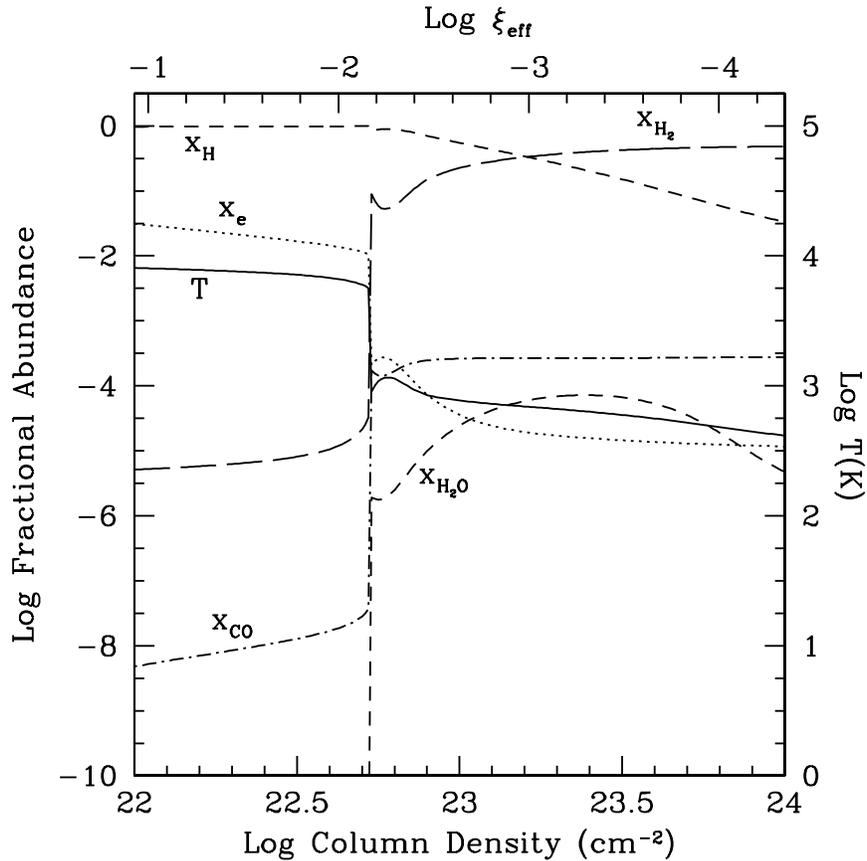}{4.4in}{0}{70}{70}{-220}{-150}
\caption{Thermal and chemical structure of an isobaric XDR. This model
assumes an X-ray source with a hard X-ray (1--100 keV) luminosity of
$10^{43}\>\ergpsec$ and an X-ray spectral index $\alpha=0.7$ at a
distance of 1 pc. The pressure has been fixed at $P/k=10^{11}\pcubcm$
K. The gas temperature, electron fraction, and the abundances of
several important species (relative to the total hydrogen density) are
plotted as a function of column density into the XDR (bottom axis) and
X-ray ionization parameter \xieff\ (top axis).}
\label{fig2}
\end{figure}

The physical conditions just past the phase transition are ideal for
producing water maser emission over a very broad range of pressures,
$P/k\approx 10^{10}-10^{13}\pcubcm$ K. Figure 3 (from NMC) shows a
more detailed look at the temperature and water abundance as a
function of depth (rather than column density) into an
X-ray-irradiated slab. Also shown is the maser volume emissivity in
saturation, \phisat. The maser volume emissivity is simply the rate at
which maser photons (in a particular transition) are generated per
unit volume; this rate can never be larger than the rate when the
maser is fully saturated (\ie when the maser radiation itself controls
the population inversion responsible for maser action; see for example
Neufeld \& Melnick 1991; Elitzur 1992). To the right of the vertical
dotted line in this panel, the maser action is quenched by radiative
trapping in the non-masing rotational transitions. However, this
calculation ignored the role of dust in the radiative transfer. As
pointed out by Collison \& Watson (1995; see also Neufeld 2000), the
presence of dust can be very important for these X-ray-powered masers,
because dust absorption sets a lower limit to the escape probability
for the far-infrared rotational transitions, and thus acts to inhibit
quenching of the maser emission, provided that the dust is
substantially cooler than the gas; otherwise the radiation from the
dust would be important. This is unique to these XDR masers, because
only there can the column densities become large enough that the
far-infrared dust optical depths become significant. The importance of
dust depends on the pressure in the XDR, and is negligible for
$P/k\gtrapprox 10^{12}\pcubcm$ K. This is because with increasing
pressure the location of the transition to the molecular regime moves
to smaller column density; since the column density at the transition
sets the column density scale on which the ionization and heating
rates decline, with increasing $P$ the XDR becomes more compressed in
column density, so the importance of dust in the radiative transfer
diminishes.

\begin{figure}
\plotfiddle{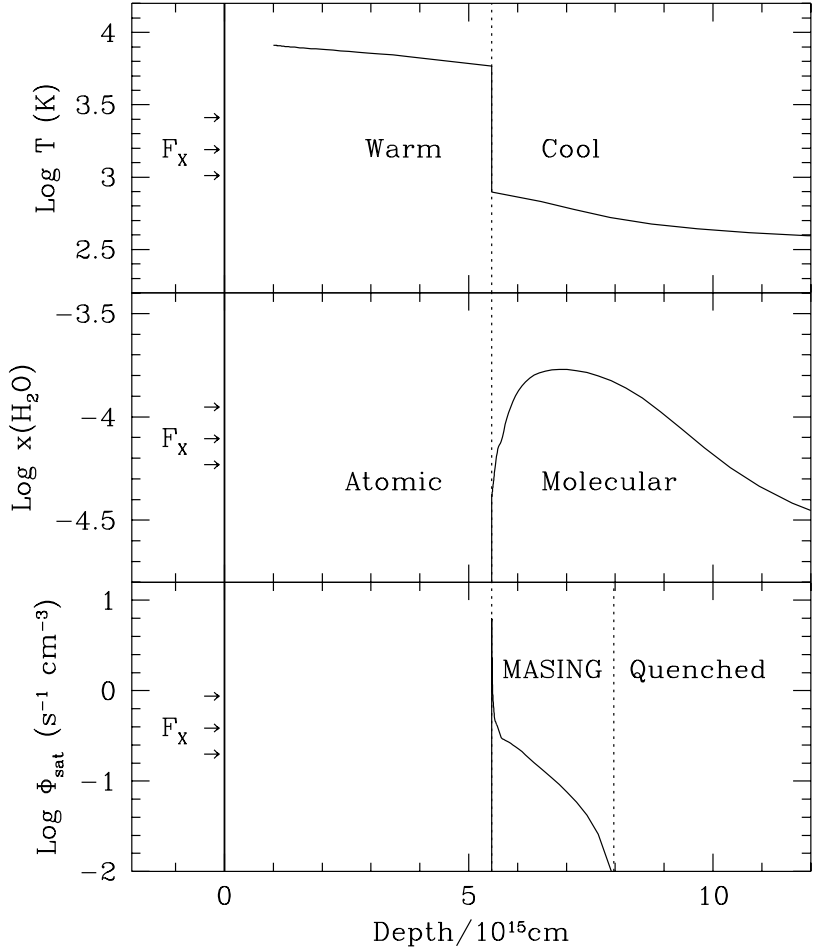}{4.8in}{0}{68}{68}{-220}{-115}
\caption{As Figure 2, except plotted as a function of depth rather
than column density. $F_x$ denotes the incident hard X-ray flux. In
addition to temperature and water abundance, the bottom panel shows
the maser emissivity \phisat\ (photons \pcubcm\ \psec); see the text
for definition. From NMC.}
\label{fig3}
\end{figure}

As shown by NMC, the resulting water maser luminosities from XDRs can
be substantial, reaching  $L_{\rm H_2O}\sim $ 10s to 100s \lsol\ per
pc$^2$ of irradiated area. Since this is the luminosity range seen in
the megamaser galaxies, with the exception of the so-called
``gigamaser'' source (Koekemoer \etal 1995), which is an order of
magnitude more luminous, it is evident that XDRs {\it can} explain the
association of \htwoo\ megamasers with AGN, provided that areas of
only $\sim 1$ pc$^2$ are being irradiated by hard X-rays from the
AGN. Since the obscuring ``tori'' inferred to be present in Type 2
Seyferts (see Antonucci 1993 for a review) were also inferred to be
$\sim $pc-scale objects, they naturally provided the requisite surface
areas to explain the megamaser luminosities, without appeal to any
special geometries.

\section{Imaging 1: NGC 4258}
Further progress in understanding \htwoo\ megamasers would depend on
imaging the maser emission. With the advent of the Very Long Baseline
Array (VLBA), it became possible to obtain very high resolution
interferometric images of the 22 GHz emission in sufficiently bright
megamaser sources. The first such galaxy to be observed was the nearby 
($D\approx 7$ Mpc), LINER spiral NGC 4258, which has a 22 GHz line
luminosity $L_{\rm H_2O}\approx 85\lsol$. Most strikingly, NGC 4258
was already known from single-dish observations (Nakai \etal 1993) to
show an enormous velocity spread in its maser emission, with 
$\Delta V$ up to $\pm 1000\kms$ with respect to the systemic
velocity. Combined with the already established concentration of the
emission to the nucleus from earlier VLA observations (Claussen \& Lo
1986), this strongly suggested that material very deep in a potential
well -- such as gas in proximity to a massive black hole -- could be
involved in the emission.

NGC 4258 was first observed using VLBI by Greenhill \etal (1995). Only
the systemic emission was imaged; nevertheless, these observations
already revealed some of the key features discussed below. Maser
emission over the full velocity range was imaged using the VLBA by
Miyoshi \etal (1995). The results, summarized in Figure 4 (from Bragg
\etal 2000), are quite remarkable. Among the notable features:
\begin{itemize}
\item{The maser emission does not arise from the inner face of an
X-ray-irradiated torus; instead, it is produced in a thin, warped
disk or annulus, with inner and outer edges at 0.14 and 0.28 pc. This
distance scale corresponds to an inner edge at 38,000 Schwarzschild
radii for the mass of the central black hole (see below). The
disk is extremely thin: the maser emission is unresolved in the
vertical direction, implying that the thickness of the masing layer is
less than $10^{15}$ cm (Moran \etal 1995).} 
\item{The rotation curve is Keplerian to better than $1\%$. This
places stringent limits on the nature of the central mass
concentration (Maoz 1995): the central mass $M=3.9\times 10^7\msol$
within 0.14 pc. This implies that the central mass density $\rho > 3.4
\times 10^9\msol$. Combined with the limits on deviation from Keplerian
motion, this rules out a stellar cluster as the central object.  If
the cluster were composed of stars or stellar remnants with masses
above $M\sim 0.03\msol$, the evaporation timescale $t_{evap}\ll t_H$,
the Hubble time. For a cluster made up of brown dwarfs or other low
mass objects with $M < 0.03\msol$, the collisional timescale
$t_{coll}\ll t_H$. In addition, a stellar density cusp around a
low-mass black hole is also ruled out. Hence NGC 4258 provides
compelling evidence for the existence of a supermassive black hole in
a galactic nucleus.}
\item{Since we know the mass of the central object and the associated
X-ray luminosity (Makishima \etal 1994) quite accurately, we can make
the first reliable determination of the fractional Eddington
luminosity in an AGN, subject only to uncertainty in the hard X-ray to
bolometric luminosity ratio. This turns out to be quite small, with
$L/L_{\rm Edd} \sim 10^{-4}$.}
\item{Observations of centripetal acceleration ($a\approx
9\kms\,\pyr$: Haschick, Baan, \& Peng 1994; Greenhill \etal 1995a;
Nakai \etal 1995) or proper motion ($\mu\approx 32$ $\mu$as \pyr:
Herrnstein 1997; Herrnstein \etal 1999) of systemic velocity features
allow a purely {\it geometric} determination of the distance to NGC
4258 ($D=7.3\pm 0.3$~Mpc); both methods give the same answer. The
tight limit on the distance quoted here comes from the data set
analyzed in the last two of these references, which contains multiple
epochs of observation from 1994 to 1997. The uncertainties are
contributed nearly equally by the statistical errors (\ie the
precision with which $a$ and $\mu$ can be determined, which is limited
mainly by the difficulty in identifying and tracking individual
features) and the uncertainties in the disk model (chiefly the
projected disk angular velocity at the radius of the systemic
features).}
\item{Observations of centripetal acceleration of the systemic and
high-velocity features (the latter having $a\lessapprox 1\kms\,\pyr$:
Bragg \etal 2000) confirm that the geometry inferred for the masers
(shown in Figure 4) is correct; the scatter of the high-velocity
features about the disk midline, as constrained by the upper limits on
$a$, is only a few degrees, while the systemic velocity features lie
within a very narrow range of radii (approximately 0.01 pc spread
about a mean radius $R=0.14$ pc). This latter conclusion is supported
by the essentially linear variation of velocity with impact parameter
$b$ for the systemic velocity features (with a gradient $dv/db\simeq
0.26\kms\;\mu{\rm as}^{-1}$); this behavior is expected as the
projection of orbital velocity onto our line of sight changes with
impact parameter (Miyoshi \etal 1995). Calculation of the implied
radius for the enclosed mass places the systemic velocity features at
the same mean radius as inferred from the centripetal accelerations.}
\item{Compact continuum emission has been detected at 22 GHz
(Herrnstein \etal 1997). The peak of the emission is located about
0.015 pc north of the dynamical center of the disk. Herrnstein \etal
suggest that this is the base of the northern jet of the symmetric
pair seen on parsec to kiloparsec scales in the radio and X-ray, and
detect a much weaker southern component, at somewhat greater distance
from the disk center, which they argue is attenuated by free-free
absorption in the warm atomic layer on that side of the disk. Assuming
the VLBI jet is intrinsically symmetrical, the failure to detect the
southern component at the same offset (0.5~mas) as the peak in the
northern jet suggests that the free-free optical depth is $\gtrapprox
2$. Herrnstein \etal argue that the systemic velocity masers are
amplifying the southern jet emission, an argument that is supported by
the correlation of the maser flux density with the continuum flux
density of the northern jet emission; this in turn implies that the
jet is symmetric, as assumed.}
\item{By searching for Zeeman-splitting-induced circular polarization
of the 22 GHz water maser emission, Herrnstein \etal (1998) derived an
upper limit of 300 mG to the magnetic field strength in the disk at a
radial distance of 0.2 pc. This upper limit is about four times larger
than the value required for an equilibrium disk supported by magnetic
pressure. However, it is still significant because one cannot rule out
{\it a priori} the possibility that the masers trace a thin layer in a
much thicker accretion disk. This upper limit to $B$ can also be used
to place an upper limit to the mass accretion rate through the disk,
under the assumption of equipartition; unfortunately, this method at
present is not restrictive enough to discriminate between competing
models for the disk (see \S 3.2).}
\item{The angle over which the maser emission from NGC 4258, which is
beamed in the plane of the disk, is detectable can be determined from
the velocity extent of the emission near the systemic velocity,
compared to the rotation velocity. This gives a beaming angle of
$8^\circ$ (Miyoshi \etal 1995). One can also estimate the observable
solid angle from the magnitude of the disk warp, which has an angular
thickness much greater than the thickness of the maser layer, and
gives a ``warp angle'' of approximately $11^\circ$. Using either
number, the observed solid angle is consistent with all Seyferts and
LINERS containing such maser disks, but only those in which our line
of sight intercepts the (nearly edge-on) disk are detectable. This
suggests that the masing regions are either identical with, or
physically associated with, the obscuring material (the
TORUS\footnote{{\bf T}hick {\bf O}bscuration {\bf R}equired by {\bf
U}nified {\bf S}chemes: see Conway 1999.}) around the AGN.}
\end{itemize}

\begin{figure}
\plotfiddle{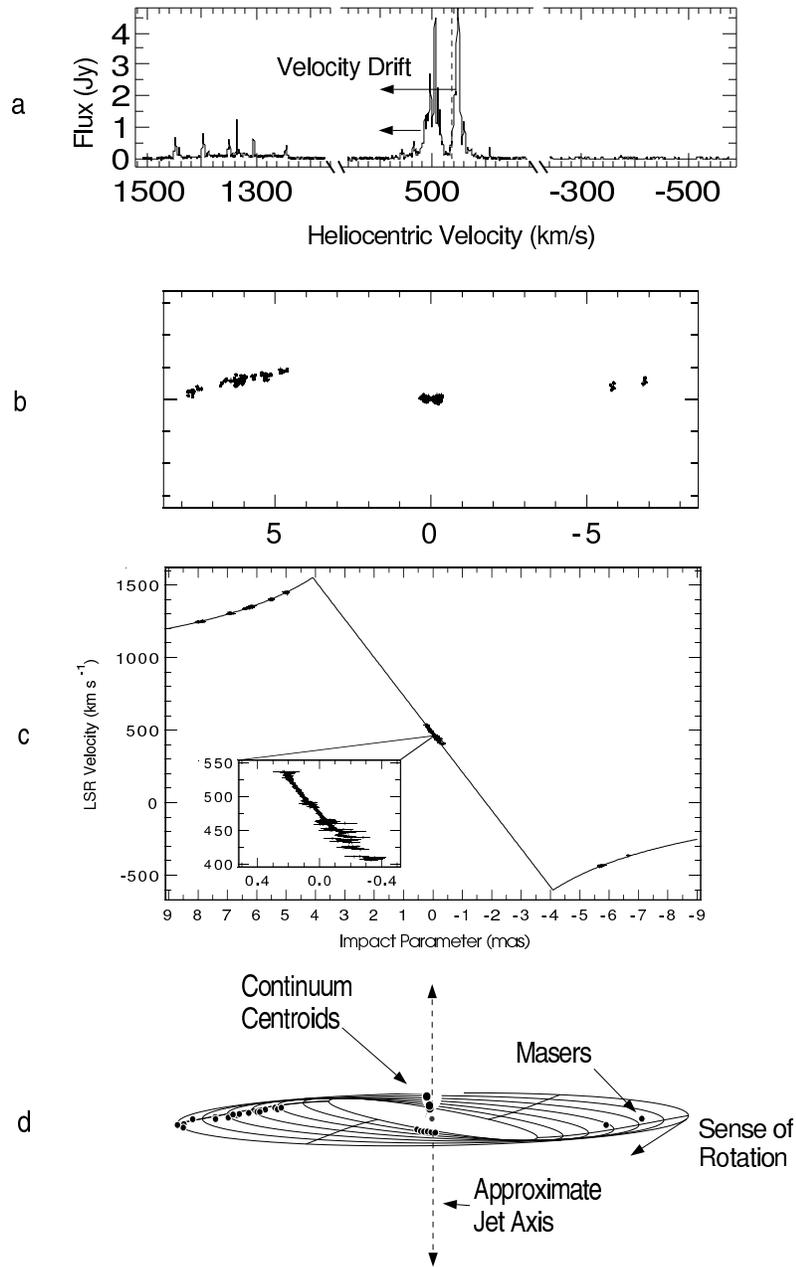}{7.2in}{0}{75}{75}{-250}{-30}
\caption{The maser disk in NGC 4248, as inferred from VLBA
observations. The top panel shows a typical spectrum of this system,
with the direction of velocity drift of the features indicated. The
second panel shows the projection of the maser spots on the plane of
the sky. The third shows the velocities of the spots as a function of
impact parameter; the solid line is a Keplerian rotation curve. The
bottom panel shows the inferred structure of the system. From Bragg
\etal 2000.} 
\label{fig4}
\end{figure}

\subsection{Deriving the Mass Accretion Rate}
The maser emission from NGC 4258 arises in what is quite clearly 
the accretion disk around the central black hole. It is perhaps
remarkable that the first direct detection of an accretion disk around
a supermassive black hole was obtained in a molecular transition,
using ground-based telescopes. We can also use the \htwoo\ maser
emission to infer the mass accretion rate through the disk (Neufeld \&
Maloney 1995; NM95). First, however, we must return briefly to the results
of NMC. 

Figure 5 shows a phase diagram for dense, X-ray-irradiated gas. The
abundance of water and the gas temperature are plotted as a function
of an X-ray ionization parameter, here written as
$F_5/(N_{24}^{0.9}\tilde P_{11})$, where the incident {\it
unattenuated} X-ray flux $F_x=10^5 F_5$ \flux, $N_{\rm att}=10^{24}
N_{24}\psqcm$ is the attenuating column density to the source of
X-rays, and the gas pressure $P=10^{11}k\tilde P_{11}$ dyne
\psqcm. (This diagram was actually calculated by varying $F_x$, but
the results will scale approximately with $N_{\rm att}$ and $P$ as
indicated.) Note that there are actually two main branches,
corresponding to the warm, atomic phase and the cooler, molecular
phase. Over a range of ionization parameter, both equilibria are
present, allowing for the possibility of a two-phase medium with the
warm atomic and cooler molecular phases coexisting. The important
point is that the upper branch in the temperature plot (the warm
atomic phase) does not exist for ionization parameters smaller than a
critical value of about 6.5. This means that for a fixed flux and
shielding column, there is a maximum pressure for which the warm
atomic phase can exist; at pressures higher than this, the gas {\it
must} be in the molecular phase.

To apply this to NGC 4258, we need a model of the accretion disk.
Since the masers appear to arise in a thin disk, model it as one: a
Shakura-Sunyaev $\alpha-$disk (Shakura \& Sunyaev 1973), but with a
warp (which is actually crucial, as shown below). In this model the
kinematic viscosity $\nu$ is assumed to be proportional to the product
of the sound speed $c_s$ and the disk scale height $H$, with the
constant of proportionality designated $\alpha$ (see Frank, King \&
Raine 1992 for a review of the theory of accretion disks). The actual
source of viscosity has been a mystery for many years but is now
generally believed to be provided by the magnetorotational instability
(Balbus \& Hawley 1991).

Because of the warp, portions of the disk see the central X-ray
source. In the obliquely illuminated region between 0.14 and 0.28
parsecs (the masing annulus), X-ray heating dominates over viscous
heating by about an order of magnitude, and the disk is nearly
isothermal. We therefore model the disk using the $\alpha$
prescription for viscosity, but the temperature is set by the X-ray
heating rather than viscous heating as in a standard $\alpha-$disk.
Ignoring self-gravity, the disk scale-height is
\be
H={c_s\over V_{\rm orb}}R = 2.6\times 10^{14} {T_3^{1/2}
R_{0.1}^{3/2}\over M_8^{1/2}}\;{\rm cm}
\ee
where the disk temperature $T=10^3 T_3$ K, the disk radius is
$R=0.1R_{0.1}$ pc, and the mass of the central black hole $M_{\rm
bh}=10^8M_8\msol$. Hence the disk is expected to be extremely thin, as
observed. The disk midplane pressure ($P/k$) is 
\be 
\tilde P = 1.3\times 10^{11} {M_8\dot M_{-5}\over \alpha R_{0.1}^3}
\;{\rm cm^{-3}\; K.}
\ee
where the mass accretion rate through the disk $\dot M=10^{-5}\dot
M_{-5}\msol\>\pyr$.
\begin{figure}
\plotfiddle{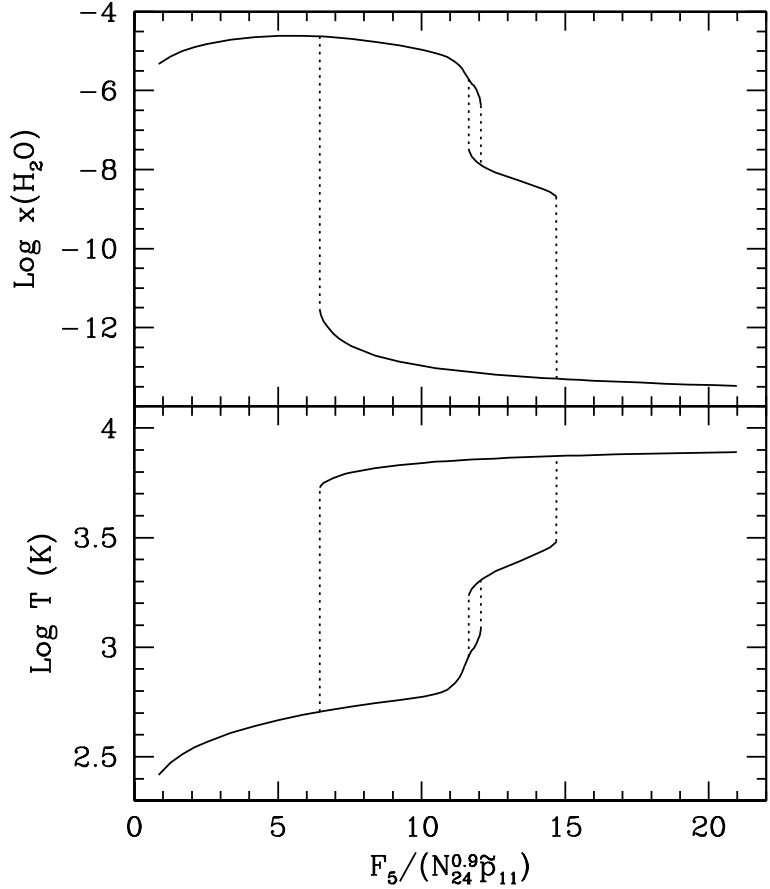}{4.8in}{0}{68}{68}{-220}{-115}
\caption{Phase diagram for dense X-ray irradiated gas. The water
abundance relative to hydrogen and the gas temperature are plotted as
a function of the X-ray ionization parameter $F_5/(N_{24}^{0.9}\tilde
P_{11})$, where the incident X-ray flux $F_x=10^5 F_5$ \flux, $N_{\rm
att}=10^{24} N_{24}\psqcm$ is the attenuating column density (\ie the
column density of gas that is neutral enough to absorb X-rays at the
energies of interest) to the
source of X-rays, and the gas pressure $P=10^{11}k\tilde P_{11}$ dyne
\psqcm. The upper and lower branches correspond to the warm atomic and
cooler molecular phases, respectively, in the bottom panel, and are
reversed in the upper panel. The existence of the intermediate
temperature phase is very sensitive to radiative trapping of cooling
lines, and it probably does not occur in real tori or disks around
AGN. Over the range of ionization parameter in which both branches
exist, a two-phase medium is possible.}
\label{fig5}
\end{figure}

If we now take the critical pressure derived from the results of NMC,
with the flux expressed in terms of the luminosity of the central
X-ray source and the disk radius, scaled to appropriate values for NGC
4258, we have that the critical pressure for molecule formation is
\be
\tilde P_{\rm cr}\approx 3.3\times 10^{10} {L_{41}\over R_{0.1}^2
N_{24}^{0.9}}\;{\rm cm^{-3}\; K.}
\ee
Equating these two expressions gives the critical radius for molecule
formation: 
\be
R_{cr}=0.04 {(\dot M_{-5}/\alpha)^{0.81} M_8^{0.62}\over L_{41}^{0.43}
\mu^{0.38}}\;{\rm pc}
\ee
where $\mu$ corrects for oblique illumination at an angle
$\cos^{-1}\mu$. In writing this expression we have also assumed
that the disk itself provides all of the shielding column density,
although violation of this assumption could only change $\dot M$ by at
most a factor of two. Notice that from equation (2), the disk midplane
pressure falls with radius as $R^{-3}$, whereas the X-ray flux from
the AGN will decrease as $R^{-2}$. This means that the X-ray
ionization parameter actually {\it increases} outwards, which means
the transition from a molecular to an atomic disk should be identified
with the {\it outer} edge of the masing annulus. Plugging in the
values appropriate for NGC 4258 ($M_8=0.39$, $L_{41}=0.4$, and
$\mu\approx 0.25$, and $R=0.28$ pc), and solving equation (4) for the
mass accretion rate, we get
\be 
{\dot M\over \alpha}\approx 7\times
10^{-5}\;\msol\;\pyr.  
\ee
With the inferred mass accretion rate, the mass of the masing disk is
only $\sim 10^3\msol$, and the neglect of the disk self-gravity is
justified.

Why is there an {\it inner} edge to the masing region? In NM95, it was
speculated that this is because the disk actually flattens out
interior to 0.14 pc, so that it is no longer illuminated by the X-ray
source. In this case the temperature will be determined by viscous
heating, and this will in general be too low to produce abundant water
or maser action (see further discussion on this point
below)\footnote{Although maser action in the 22 GHz line requires that
the gas density be less than $n\sim 10^{10}\pcubcm$, it is unlikely
that the inner edge is produced by the increase of the density with
decreasing radius. As the midplane density rises, the masing layer
will simply move to a greater height above the disk. See also the
discussion in \S 3.2}. Such a disk geometry is expected in the case of
radiation-driven warping (Pringle 1996; Maloney, Begelman, \& Pringle
1996) since in this case the warp grows from the outside
inwards. Whatever the mechanism producing the warp, its presence is
crucial in the X-ray-powered maser model, since it is the warp that
allows the central source to irradiate the disk and produce the maser
emission.

The physical conditions in the disk inferred in this model are shown
in Figure 6. Beyond the critical radius at 0.28 pc, the disk is forced
by the X-ray irradiation into the warm atomic state; the temperature
is about 8000 K, and the scale height is about $H\sim 5\times 10^{15}$
cm. Molecular abundances are negligible. At the critical radius, the
pressure is just high enough at the midplane to force the gas into the
molecular phase. With decreasing radius the thickness of the molecular
zone increases as a larger fraction of the disk is above the critical
pressure; the temperature in the molecular zone will rise with
increasing $z-$height above the midplane. Interior to 0.14 pc, where
the warp flattens out, the disk temperature drops below 100 K, and
masing ceases.

\begin{figure}
\plotfiddle{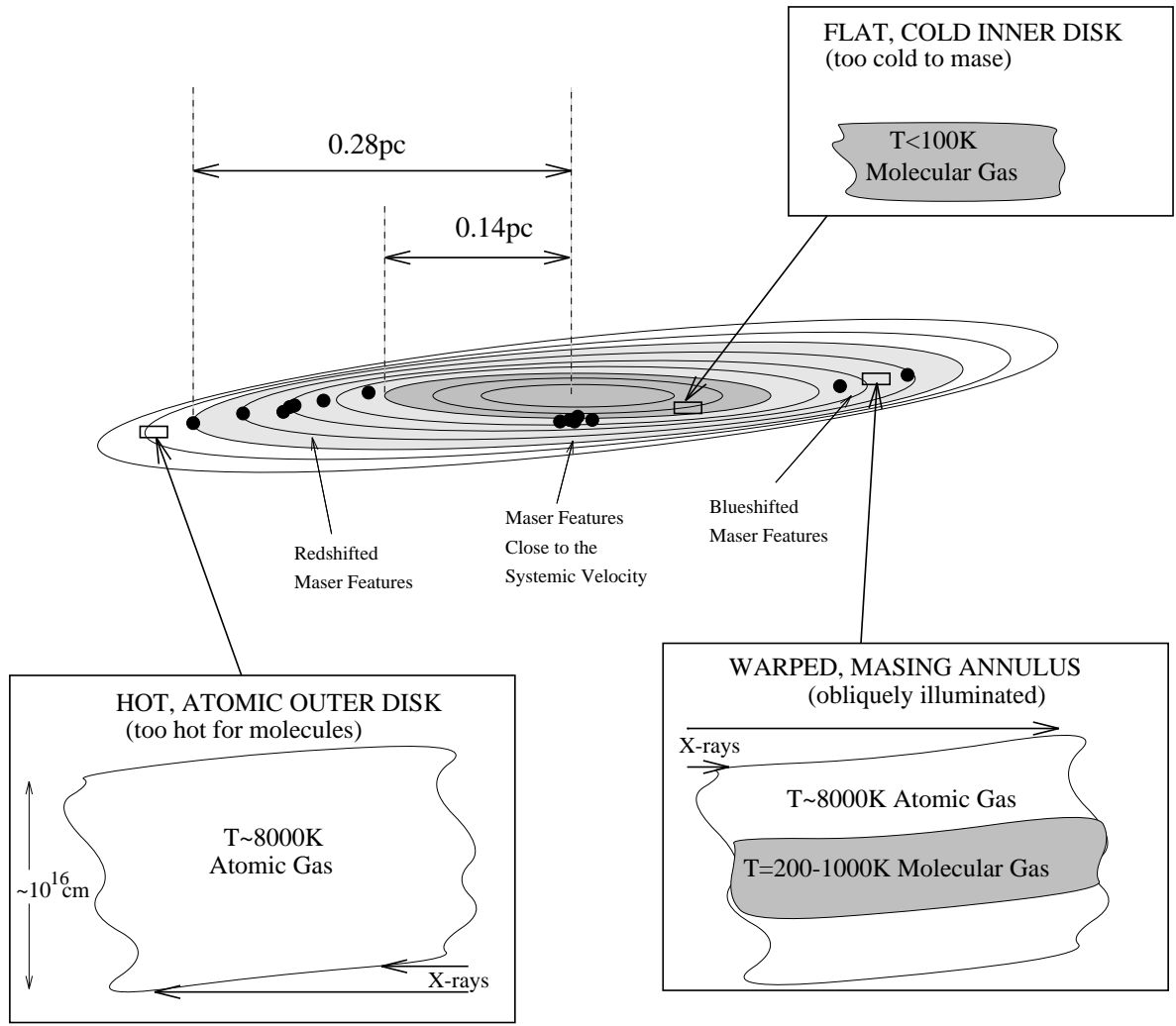}{4.7in}{-90}{68}{68}{-270}{430}
\caption{Structure of the accretion disk in NGC 4258, as inferred from
observations of water maser emission. The disk temperature within the
masing region is set by the X-ray heating rate; the outer edge is
determined by the disk pressure (where it undergoes a phase transition 
from partly molecular to completely atomic) and the inner edge results
from the flattening of the disk warp, so that it is no longer
irradiated by the central source. From Neufeld \& Maloney (1995).} 
\label{fig6}
\end{figure}

If we assume that $L_x/L_{bol}\sim 0.1$, as is typical for AGN, the
derived mass accretion rate implies that rest-mass energy is converted
to radiation with a radiative efficiency 
\be
\epsilon \sim 0.1/\alpha
\ee
(assuming steady accretion over the time it takes material to reach
the black hole from 0.28 pc; the validity of this assumption is rather
uncertain, as discussed briefly below). This in turn indicates that the low
fractional Eddington luminosity of NGC 4258 is due to the low mass
accretion rate onto the central black hole, and {\it not} due to low
radiative efficiency of the accreting material. Also, although we do
not know {\it a priori} where our line of sight to the X-ray source
intercepts the disk, the column densities inferred for the disk
strongly suggest that our sightline passes through the disk just at
the edge of the molecular zone (NM95). 

\subsection{Alternative models}
The model of Neufeld \& Maloney (1995), as outlined above, is
essentially the ``what-you-see-is-what-you-get'' model: since the
maser emission appears to arise in a thin, warped, disk, assume that
this is in fact the case, with the warping of the disk allowing these
regions of the disk to see the central X-ray source, which then powers
the maser emission by a well-understood process (NMC). Unsurprisingly,
in view of the uniquely detailed information on an accretion disk
around a supermassive black hole provided by the maser observations, a
number of additional models have been proposed. These will be briefly
discussed here.

Maoz \& McKee (1998) (MM98) proposed that rather than X-ray heating,
the maser emission in NGC 4258 is due to shocks in spiral density
waves in a self-gravitating disk\footnote{Note that in any scenario in
which the maser emission arises in a disk, we must always view it
close to edge-on if we are to see the maser emission, which will be
strongly beamed in the disk plane.}. The main motivation for this was
the hint of ``regular'' spacing in radius and velocity of the
high-velocity features in the disk, as seen in the data of Miyoshi
\etal (1995) (see Figures 4b and c). Although this model contains some
{\it ad hoc} assumptions (\eg it is assumed that conditions in the
disk are such that water maser action will occur when a density wave
-- which by fiat produces a shock of the required speed -- passes
through the disk between 0.14 and 0.28 pc, and nowhere else; also, all
of the predicted maser luminosities are scaled to a disk thickness of
$10^{16}$ cm, an order of magnitude larger than the observed upper
limit to the dimensions of the maser spots), it did make very definite
predictions about the pattern of accelerations that should be seen as
a function of impact parameter. This model also explained the relative
weakness of the blue-shifted high-velocity features relative to the
red-shifted features as a consequence of absorption by non-masing
water. Indeed, Maoz \& McKee made a generic prediction that all such
disks should exhibit the same blue/red asymmetries, since this is an
intrinsic feature of the spiral shock model. The mass accretion rate
in this model is approximately two orders of magnitude larger than the
rate inferred by NM95.

The predicted acceleration pattern was tested by Bragg \etal (2001),
who found no evidence to support the MM98 model, for any
chosen pitch angle of the spiral pattern. The key point is that in the
MM98 model, all of the blue-shifted features should exhibit positive
accelerations and the red-shifted negative, but in fact positive and
negative $\dot v$ are seen for both sides. Furthermore, both NGC 3079
(Nakai \etal 1995) and NGC 5793 (Hagiwara \etal 1997) have exhibited
blue-shifted emission that is stronger than the red-shifted emission
(always, in the case of NGC 3079), while the high-velocity emission
features in IC 2560 are nearly symmetric in intensity (Ishihara \etal
2001). Hence at present there is no reason to suppose that this model
applies to NGC 4258, or any other megamaser source.

Desch, Wallin, \& Watson (1998) proposed that rather than X-ray
irradiation, the maser emission in NGC 4258 is powered by viscous
dissipation within the accretion disk itself. This requires that the
energy generated by viscous dissipation is deposited in a manner which
is not proportional to the local density. The viscous heating rate per
unit disk face area is 
\be
D(R)={3GM\dot M\over 8\pi R^3}
\ee
while the disk surface density 
\be
\Sigma(R)={\dot M\over 3\pi\nu}
\ee
where the radial dependence of $\Sigma$ is contained entirely in
$\nu$. Both equations (7) and (8) assume a steady-state disk far from
the inner edge. Doubling equation (7) to get the total heating rate
through the disk, the ratio of viscous heating rate to surface density
is
\be
{2D(R)\over \Sigma(R)}={9\over 4} {GM\nu\over R^3}
\ee
which is {\it independent} of $\dot M$. This is also the
vertically-averaged heating rate per particle, which,
again assuming the $\alpha-$prescription for viscosity and a thin,
Keplerian disk, as above, can be written as
\be
{\bar\Gamma(R)\over \bar n_{\rm
H_2}(R)}=2.1\times 10^{-24}{\alpha M_8^{1/2}c_1^2\over R_{\rm
pc}^{3/2}}{\;\rm erg\ s^{-1}}
\ee
where $\bar\Gamma(R)$ is the vertically-averaged heating rate per
unit volume and $\bar n_{\rm H_2}$ is the similarly averaged number
density of molecular hydrogen; $c_1$ is the gas sound speed in units
of 1 \kms\ (appropriate for molecular gas at a few hundred K). Note
that this value of the viscous heating rate per particle is
independent of how $\bar\Gamma$ and $\bar n_{\rm H_2}$ are calculated
from $D$ and $N_{\rm H_2}$, respectively, provided only that they are
calculated the same way. Plugging in the numbers appropriate to NGC
4258, 
\be
{\bar\Gamma\over \bar n_{\rm H_2}}=3.8\times10^{-23}\alpha \;{\rm
erg\ s^{-1}.}
\ee
For $\alpha\lessapprox 1$, this is approximately two orders of magnitude
smaller than the heating rate per particle needed to power \htwoo\
maser action in the disk (Desch \etal 1998). Hence it is necessary to
assume that the viscous energy is deposited in a different fashion
with height $z$ than the gas density $\rho(z)$, a not implausible
assumption. 

The simplest assumption (and one that is in reasonable agreement with
the results of Stone \etal 1996, who explicitly calculated the energy
deposition as a function of height in three-dimensional
magnetohydrodynamic simulations of the magnetorotational instability)
is to assume that the viscous heating rate per unit volume is
independent of $z$. In this case the heating rate per particle simply
increases with height as a Gaussian with scale height equal to the gas
scale height $H$; the ratio of the heating rate per particle to the
vertically-averaged heating rate per particle derived above (equation
9 or 10) is 
\be {\Gamma/n(z)\over \bar\Gamma/\bar n}\simeq
\exp(z^2/2H^2) 
\ee 
which is (not surprisingly) independent of everything except the
height above the disk midplane in scale heights. In particular, this
is independent of $\dot M$; it is just the ratio of the midplane
density to the density at height $z$.

To maintain high enough temperatures to power the maser emission, we
require a local heating rate which exceeds the vertically-averaged
rate by a factor of about $100/\alpha$. Thus it is necessary that
\be
\exp(z^2/2H^2)\approx 100/\alpha \qquad\Longrightarrow \qquad {z\over
H}\gtrapprox 3,
\ee
even for $\alpha\approx 1$, {\it i.e.}, the density must drop to
$1/100$ of the midplane value. The only constraints on $\dot M$ are
then:
\begin{itemize}
\item[(a)] The densities three or more scale heights above the plane must
be high enough to excite the masing transitions;
\item[(b)] The column density of masing gas must be large enough to be
detectable.
\end{itemize}

This can be made to work for NGC 4258, for values of $\dot M/\alpha$
comparable to (or larger than) the NM95 rate, in the sense that the
above conditions can be satisfied. While the outer edge can
be understood as the result of either the gas density or column
density declining below the threshold for maser emission, the inner
radial cutoff is unexplained, since the warping of the disk is
irrelevant, and the masing layer should simply move higher into the
disk atmosphere with decreasing radius. However, it appears very
unlikely that this mechanism can apply to the other masing disk
systems that have been observed (see section 4). In general, the
factor by which the local viscous heating rate (in the masing region)
has to exceed the vertically-averaged rate is
\be
{\Gamma/n(z)\over \bar\Gamma/\bar n}\sim 1300 {R_{\rm pc}^{3/2}\over
\alpha M_8^{1/2}}.
\ee
For NGC 1068, for example, with $R_{\rm pc}\sim 0.65$, $M_8\sim 0.15$,
this is a factor of $\sim 1800/\alpha$. For the other systems that
appear to be disk-like (Moran \etal 1999; \S4) the spatial scales are
also typically $r_{\rm maser}\sim $ 1 pc, while the masses of the
central objects are five to ten times smaller than in NGC 4258 or NGC
1068. This leads to prohibitively large values of $\dot M/\alpha$, \eg
in excess of 2 \msol\ \pyr\ for the maser disk in NGC 1068. This is
inconsistent with the mass and luminosity of the central source in NGC
1068 (which, radiating at nearly Eddington luminosity, is most
definitely {\it not} an ADAF: see below), unless $\alpha\sim 10^{-2}$.
However, since the value by which the local heating rate must exceed
the vertically-averaged value itself scales as $\alpha^{-1}$ (equation
14), the discrepancy cannot be solved this way, since the values of
$\dot M/\alpha$ required to satisfy the density or column density
requirements will simply scale up proportionately as $\alpha$ is
decreased. Hence I am rather skeptical about the viability of this
mechanism as a general explanation for water maser emission in
AGN\footnote{There is an additional aspect of this model that is
rather problematic for high accretion rates. As noted earlier, a
fundamental requirement for maser action is departure from
thermodynamic equilibrium, in particular, that the gas temperature
exceed the temperature of the radiation field in the mid- to
far-IR. Desch \etal calculated the temperatures in their model under
the assumption that the dust temperature is determined by the
continuum emission of the central source incident on the disk, while
the gas temperature is determined by viscous heating, and conclude
that $T_{\rm gas}$ can exceed $T_{\rm dust}$ by a large enough factor
to allow maser emission. This may be plausible for the NM95 mass
accretion rate, since in this case the disk is not very optically
thick (at least in the masing region) in a Rosseland mean sense, and
the local viscous heating is substantially smaller than the incident
continuum. For accretion rates that are much larger than this,
however, neither of these statements are true, and it is difficult to
see why the dust and gas temperatures will not be well coupled in a
disk heated dominantly by viscous heating.}.

Kartje, K\"onigl, \& Elitzur (1999) proposed a variant X-ray powered
accretion disk model, which differs from NM95 in (a) being
specifically clumpy (although the NM95 model is referred to throughout
as the ``homogeneous disk model'', as though this is a
requirement, the same arguments carry over to a clumpy disk,
the only difference being that the critical pressure refers to the
clumps containing the bulk of the mass at a given radius); (b) has the
clumps in NGC 4258 occur in a disk-driven hydromagnetic wind: the disk
itself is actually flat, not warped; and (c) has a mass accretion rate
that is two orders of magnitude higher than that derived by NM95, similar
to MM98. One might perhaps be somewhat surprised that the observed
disk should appear so thin, warped, and Keplerian if the emission
actually arises in clumps uplifted off the flat disk in a wind.

Since the NGC 4258 portion of this model has parameters that have of
course been chosen to match the observations, perhaps its most direct
prediction is that the mass accretion rate is much larger than
inferred by NM95, exceeding $\dot M\approx 10^{-2}\msol\pyr$. This is
also true of the MM98 model, since accretion rates much higher than
the NM95 rate are required to make the disk self-gravitating, and high
mass accretion rates were also favored by Desch \etal (1998). Since
the luminosity of the nucleus of NGC 4258 is well determined
observationally (Makishima \etal 1994), all of these high mass
accretion rate models require that the inner portion of the accretion
disk (assuming steady accretion; I will return to this point shortly)
radiates very inefficiently. Hence these papers have all appealed to
``ADAF'' models (for Advection-Dominated Accretion Flow; this is
really just a renaming of what used to be referred to as ``ion tori''
[See Svensson 1999]) for accretion, in which the inner region of the
flow is very hot (with the ions at close to the virial temperature),
but radiates very inefficiently, as the electrons are not well coupled
to the protons (so that $T_e \ll T_i$) (Ichimaru 1977; Rees \etal
1982; Phinney 1983; the modern incarnations begin with Narayan \& Yi
1994 and Abramowicz \etal 1995, with literally dozens of follow-up
papers). These flows exist only for values of $\dot M$ below a
critical threshold; for higher $\dot M$ only the standard thin,
``cold'' (\ie $T \ll T_{\rm vir}$ disk solutions exist.

Indeed, given the extremely low fractional Eddington luminosity of NGC
4258, there were immediate attempts to fit it into the ADAF framework
(Lasota \etal 1996), since if its accretion is not described by an
ADAF, there is really no reason to assume that ADAF models apply to
any AGN. (As noted above, the fractional Eddington luminosity for NGC
1068 is so high -- nearly unity -- that there is no question that it
is {\it not} described by an ADAF solution.) The success of the ADAF
model lies in the fits to the ``spectrum'' of NGC 4258 (for an
example, see Figure 1 of Lasota \etal 1996). If $\dot M$ is in fact
much larger than the NM95 rate, we must conclude that the disk does
not see the central X-ray source beyond 0.28 pc, since otherwise maser
emission would arise further out, since there is no longer a
transition to an atomic disk at this radius, as $R_{cr}$ will also be
much larger (see equation [4]).

At present the ADAF models for NGC 4258 have been confined to an
increasingly small portion of parameter space; the absence of 22 GHz
continuum emission associated with the ADAF (Herrnstein \etal 1998)
and the detection of X-ray variability on timescales of $\sim
4000-40,000$ seconds (Fiore \etal 2001) indicate that the transition
to an ADAF must occur on scales of order 100 Schwarzschild radii or
less. The most definitive test for the presence of cold material close
to the central black hole would be the detection of a broad Fe
K$\alpha$ line; there has not yet been an observation of sufficient
sensitivity (Reynolds, Nowak, and Maloney 2000; Fiore \etal 2001) to
either detect such a line or rule one out at an interesting level.

One important point that has been raised in connection with the mass
accretion rate and the radiative efficiency is the assumption of
steady-state accretion (Gammie, Blandford, \& Narayan 1999). As noted
earlier, the efficiency inferred by NM95 is only valid if accretion
has been steady over the time it takes material to migrate from the
radius of the maser disk down to the central source. However, if the
disk flattens out at the inner radius of the maser disk, the viscous
timescale increases by a large factor due to the drop in disk
temperature, and exceeds $10^9$ years. It is very unlikely that
accretion has been steady on this timescale. However, viscous heating
at this radius will only heat the disk to $T < 100$ K (NM95). This
means that the disk will be marginally self-gravitating: the Toomre
$Q$ parameter
\be
Q={3 c_s^3\alpha \over G \dot M}\simeq 4.0\times 10^{-3}
T_3^{3/2} {\alpha\over \dot M}
\ee
is $< 1.8$ for the NM95 mass accretion rate, at these low
temperatures. It therefore seems plausible that the mass flow through
the disk at these radii will be determined by gravitational
instabilities rather than viscosity, and so the magnitude of the
viscous timescale may be irrelevant. 

\section{Imaging 2: Is NGC 4258 Unique?}
NGC 4258 was the first of the megamaser sources to be imaged with
VLBI. The remarkable results obtained from this experiment generated a
great deal of interest in imaging of the other \htwoo\ megamaser
sources. With existing instrumentation it is difficult to obtain VLBI
observations of masers that are weaker than about 0.5 Jy because of
the necessity of detecting the maser within the coherence time of the
interferometer (Moran \etal 1999). By using a maser feature as phase
reference, and referring the other maser features to it, it is
possible to extend the coherence time, and thereby observe weaker
features, but it is still a difficult endeavour (J.~Moran,
pers. comm.) Hence not all of the known sources can be imaged at
present. To date, nine megamasers have been imaged with VLBI. Of these
nine, four show strong evidence for disk structure and two show
probable evidence for disks from the spatial and velocity distribution
of the emission; one, for which only the systemic emission has yet
been imaged, shows kinematic evidence for a disk and may be the most
similar to NGC 4258 of any source yet observed. The other two sources
appear to be very different in origin. Table 1 (adapted from Moran
\etal 1999) summarizes the results. None of the disk sources are as
well-defined as in NGC 4258, nor do any of them exhibit Keplerian
rotation curves. This latter feature, at least, is not unexpected, at
least in the X-ray powered maser model, as I discuss below. First I
will review some of the more interesting results from the imaging
studies.

\begin{table}[b]
\begin{center}
\begin{tabular}{l@{}ccrcccl}
\multicolumn{8}{c}{TABLE 1$^a$}\\[1ex]
\multicolumn{8}{c}{\sc Water Megamasers Observed with VLBI}\\[1ex]
\hline\hline\noalign{\vspace{2ex}}
\multicolumn{8}{c}{Masers With Disk Structure}\\[1ex]
\hline\noalign{\vspace{1ex}}
     & $D$ & $v_{\phi}$ & \omit{\hfil $R_i$/$R_o$\hfil} & $M$ & $\rho$ & $L_x$ &        \\
Galaxy& \sm Mpc & \sm km/s & \omit{\hfil \sm pc\hfil} & \sm 10$^6$ M$_{\odot}$
 & \sm  10$^7$ M$_{\odot}$/pc$^3$ & \sm 10$^{42}$ erg/s & Reference \\ 
\hline\noalign{\vspace{1ex}}
NGC 4258 & \z 7 & 1100    & 0.13/0.26   &   39 & 400 & 0.04 & M95\\
NGC 1068 & 15   & \z  330 & 0.6/1.2\z   &   17 &   3 & 40   & GG97\\
Circinus & \z 4 & \z  230 & 0.08/0.8\z  & \z 1 &  40 & 40   & G01\\
NGC 4945 & \z 4 & \z  150 & 0.2/0.4\z   & \z 1 &   2 &  6   & G97\\
NGC 1386 & 12   & \z  100 & --/0.7\z    & \z 2 &   4 & 0.02 & B99\\
NGC 3079 & 16   & \z  150 & --/1.0\z    & \z 1 & 0.2 & 1--10 & T98, S00\\
IC 2560$^b$ & 26   & \z  420 & 0.07/0.26\z  & \z 3 & 210 & 0.1 &  I01\\
\noalign{\vspace{.5ex}}\hline
\hline
\noalign{\vspace{2ex}}
\multicolumn{8}{c}{Masers Without Obvious Disk Structure}\\[1ex]
\hline\noalign{\vspace{1ex}}
     && $D$     & \omit{\hfil $v_0$\hfil}  & $\Delta v$ & $\Delta R$ &         &           \\
Galaxy&& \sm Mpc & \omit{\hfil \sm km/s\hfil}  & \sm km/s   & \sm pc     & Comment & Reference \\
\hline\noalign{\vspace{1ex}}
IRAS 22265 (S0) && 100 & \omit{\hfil 7570\hfil} & 150   & 2.4 & messy      & G02\\
NGC 1052 (E4)   && \z 20  & \omit{\hfil 1490\hfil} & 100   & 0.06 &
``jet''    & C98\\
\noalign{\vspace{0.5ex}}
\hline
\noalign{\vspace{1.5ex}}
\multicolumn{8}{l}{\parbox{5.8 in}{\parindent=0pt \baselineskip=10pt
\raggedright
$^a$Adapted and updated from Moran, Greenhill, \& Herrnstein (1999)}}\\
\noalign{\vspace{1ex}}
\multicolumn{8}{l}{\parbox{5.8 in}{\parindent=0pt \baselineskip=10pt \raggedright
$D$ = distance,
$v_0$ = systemic velocity, $\Delta v$ = velocity range, $\Delta R$ = linear extent,
$v_{\phi}$ = rotational velocity, $R_i$/$R_o$ = inner/outer radius of disk, 
$M$ = central mass, $\rho$ = central mass density, $L_x$ = X-ray luminosity.}}\\
\noalign{\vspace{1ex}}
\multicolumn{8}{l}{\parbox{5.8 in}{\parindent=0pt \baselineskip=10pt
\raggedright
$^b$Only the systemic velocity emission in IC 2560 has been imaged;
it is included here because of kinematic evidence for a rotating
disk (Ishihara \etal 2001). See discussion in text.}}\\
\noalign{\vspace{1ex}}
\multicolumn{8}{l}{\parbox{5.8 in}{\parindent=0pt \baselineskip=10pt
\raggedright
References: G02 = Greenhill \etal 2002, C98 = Claussen \etal 1998,
I01 = Ishihara \etal 2001, M95 = Miyoshi \etal 1995, GG97 = Greenhill
\& Gwinn 1997, G01 =Greenhill \etal 2001, G97 = Greenhill \etal
1997, B99 = Braatz \etal 1999, T98 = Trotter \etal 1998, S00 =
Sawada-Satoh \etal 2000}}\\ 

\end{tabular}
\end{center}
\end{table}

\subsection{NGC 1068}
The archetypal Seyfert 2 galaxy NGC 1068 was the second \htwoo\
megamaser source to be imaged with VLBI. Single-dish observations had
already shown that maser emission extended up to about $\pm 300\kms$
away from the systemic velocity; unlike NGC 4258, the systemic
emission is not the strongest. Re-analysis of archival VLA
observations by Gallimore \etal (1996) demonstrated that, in addition
to the nuclear masers (\ie the masers associated with the radio
continuum emission believed to be coincident with the nucleus), a
second group of fainter maser features was located about $0.3''$ (30
pc) away, at the same position angle as the radio jet. The initial
VLBA observations (Greenhill \etal 1996) covered only the red and
systemic velocities, and revealed a linear structure extended over
about 1 parsec, oriented at 45$^\circ$ to the radio jet axis. Velocity
gradients were suggestive of a rotating structure. Greenhill \etal
suggested that this was the upper surface of an irradiated torus, and
predicted that the blueshifted emission, when imaged, would form the
mirror image to the redshifted. The lower (southern) half, which was
missing, might be hidden by free-free absorption.

However, when the blueshifted emission was imaged, it appeared as a
continuation of the redshifted emission, so that the maser emission
forms one linear structure that is highly inclined to the radio jet
axis. This can be fitted by a thin, rotating disk, with rotation
velocities up to $330\kms$ (Greenhill \& Gwinn 1997). The fall-off of
rotation velocity with radius is slower than Keplerian, however,
declining approximately as $r^{-0.35}$. This could be due to either
the mass of the disk itself, or to a central stellar cluster. There is
a substantial spread of the velocities about the rotation curve,
perhaps indicating a velocity dispersion of some tens of \kms. The
scale of the masing region is larger than in NGC 4258, with inner and
outer edges at approximately $0.65$ and $1.1$~pc, respectively. The
enclosed mass is about $M=1.5\times 10^7\msol$; nearly all of this
must be contributed by the black hole.

The orientation of the disk with respect to the radio jet axis appears
peculiar. However, Gallimore, Baum, \& O'Dea (1997) imaged what
appears to be a thermal source, with brightness temperatures $T_b\sim
10^5-10^6$ K between 5 and 8 GHz, that lies interior to the maser disk
and is elongated perpendicular to the jet axis. This appears to be
emission from the disk rather than the jet. If this is the case,
then interpreting the thermal source and masing disk as one continuous
structure would imply that the masing region in fact traces a severely
warped disk, with a much greater inclination warp than seen in NGC
4258. Support for a disk geometry, warped or not, is also provided by
the absence of accelerations of the high-velocity features: Gallimore
\etal (2001) derive upper limits for the redshifted masers of $<
0.01\kms\;\pyr$ from fifteen years of monitoring data with the
Effelsberg 100m telescope.

Gallimore \etal also argue that the nuclear masers vary coherently on
timescales of months to years, which is much shorter than the
dynamical timescale, and that therefore they must be responding to
variations of the central power source. Between October and November
of 1997 there was a simultaneous flare of blueshifted and redshifted
maser features, which can be naturally understood as reverberation in
a rotating disk. Neufeld (2000) has shown that the masers can respond
to variations in the central source luminosity in two ways: increasing
as the X-ray luminosity increases, which raises the maser emissivity,
or decreasing as the bolometric luminosity increases, which decreases
the difference between the dust and gas temperatures and therefore
reduces the maser output. Either mechanism can explain the variation
reported by Gallimore et al., provided that the density in the masing
regions is at least $n_\htwo\sim 10^8\pcubcm$, which is required to
obtain a variability timescale as short as seen for the maser
flare. Since, unfortunately, we do not have a direct view of the X-ray
source in this galaxy, but see it only in reflection (as is also true
for NGC 3079; see \S 4.3), it is not clear
which mechanism is actually at work.

As noted above, NGC 1068 also possesses fainter maser emission,
located approximately 30 pc away from the nucleus along the radio jet
axis (Gallimore \etal 1996). Given their location, these masers may
arise in a shock that arises at the interface between the radio jet
and a near-nuclear molecular cloud. In support of this interpretation,
the jet appears to be diverted by its interaction with the ISM at this
location.

\begin{figure}
\plotfiddle{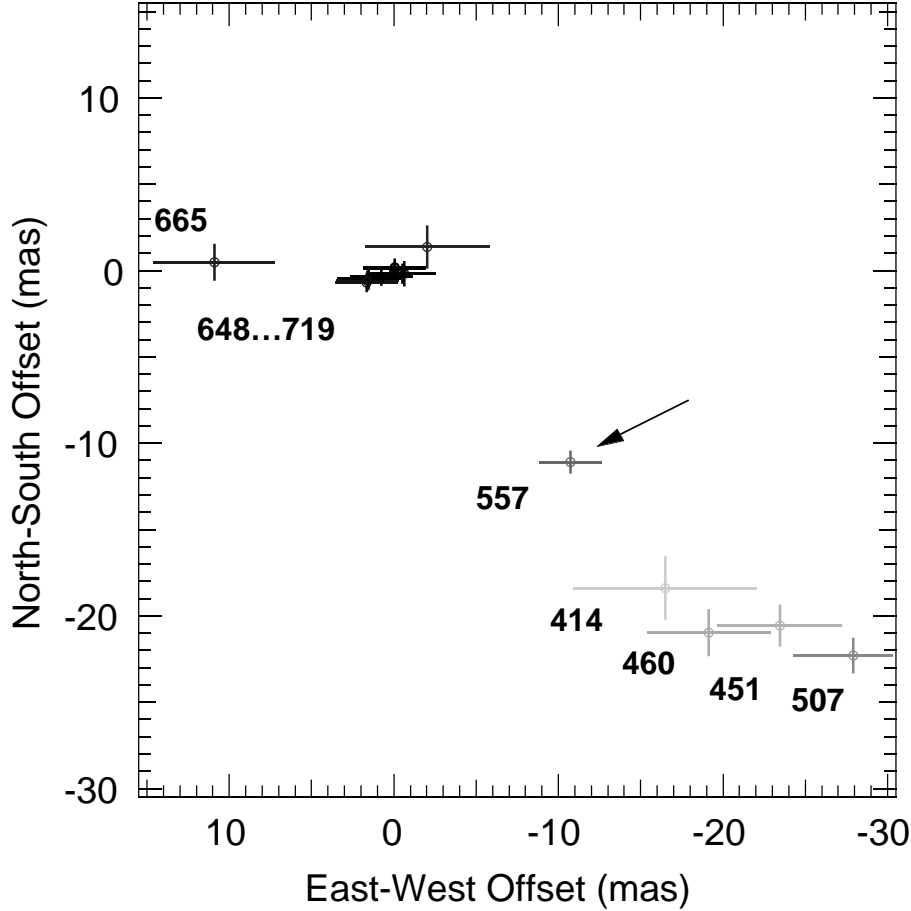}{4.5in}{0}{100}{100}{-325}{-250}
\caption{The spatial distribution of \htwoo\ maser emission from NGC
4945. The labels indicate the velocities of the emission, which is
also shown by the grayscale (lighter shading corresponds to bluer 
velocity). The arrow marks the position of the maser feature closest
to the systemic velocity of the galaxy ($V_{\rm sys}=561\kms$). At the
distance of NGC 4945, $1\ {\rm mas} = 1.9\times 10^{-2}$ pc. 
From Greenhill \etal 1997a.} 
\label{fig7}
\end{figure}

\subsection{NGC 4945}
The first \htwoo\ megamaser to be discovered (dos Santos \& Lepine
1979), NGC 4945 contains a luminous ($L_x\approx 6\times
10^{42}\>\ergpsec$) nuclear hard X-ray (1--100 keV) source (Iwasawa \etal
1993; Done, Madejski, \& Smith 1996; Madejski \etal 2000), which is
heavily obscured behind a column $N_H\approx 5\times 10^{24}\psqcm$,
so that the direct X-ray emission is only observable above $E\approx
20$ keV. In fact, the obscuring column is so high that a proper
treatment of the radiative transfer to take the substantial Thomson
optical depth into account is necessary; this raises the hard X-ray
luminosity by about an order of magnitude over simple obscuring column
density models: see Madejski \etal 2000. The X-ray emission is also
highly variable, with $t_{\rm var}\sim 1$ day, implying that the sizescale
of the emitting region is no more than $r\sim 10^{15}$ cm, and that
the absorbing material cannot occupy a large solid angle around the
disk (Madejski \etal 2000). The nuclear far-infrared luminosity is
approximately an order of magnitude larger than the hard X-ray
luminosity, and arises nearly entirely from a region no more than 225
pc by 170 pc in size (Brock \etal 1988).

NGC 4945 lies in the southern hemisphere ($\delta=-49^\circ$), and it
is therefore very difficult to observe with the VLBA. Nevertheless,
Greenhill, Moran, \& Herrnstein (1997) succeeded in observing the
maser emission using two of the southernmost VLBA antennas. A typical
spectrum of the emission (which is time variable) is shown in Figure
1. The high velocity emission brackets the systemic emission, with
velocities up to about $\pm 150\kms$ away from systemic; as in NGC
4258, the redshifted emission is much stronger than the blueshifted,
but as in NGC 1068, the redshifted emission is also much stronger than
the systemic. The spatial distribution is shown in Figure 7; note that
not all of the detected emission, particularly near the systemic and
blueshifted velocities, could be mapped. The structure is quite
linear, tracing a shallow $\cal{S}$ shape on the sky, and extends over
about 40 mas (0.8 pc). The position-velocity diagram (Figure 8) bears
some similarity to that of NGC 4258; however, the redshifted emission
does {\it not} mirror the blueshifted emission, and the fall-off of
velocity with impact parameter on the blueshifted side appears to be
faster than Keplerian, which would imply that these features do not
all lie in the plane of the sky. If one, nevertheless, interprets the
emission in terms of an edge-on, rotating disk, the implied mass is
$M\sim 1.4\times 10^6\msol$, and the central mass density is at least
$\rho\sim 2\times 10^7\msol\pcubpc$. With this central mass and the
observed luminosity, NGC 4945 must be radiating in excess of $10\%$ of
the Eddington limit.

\begin{figure}
\plotfiddle{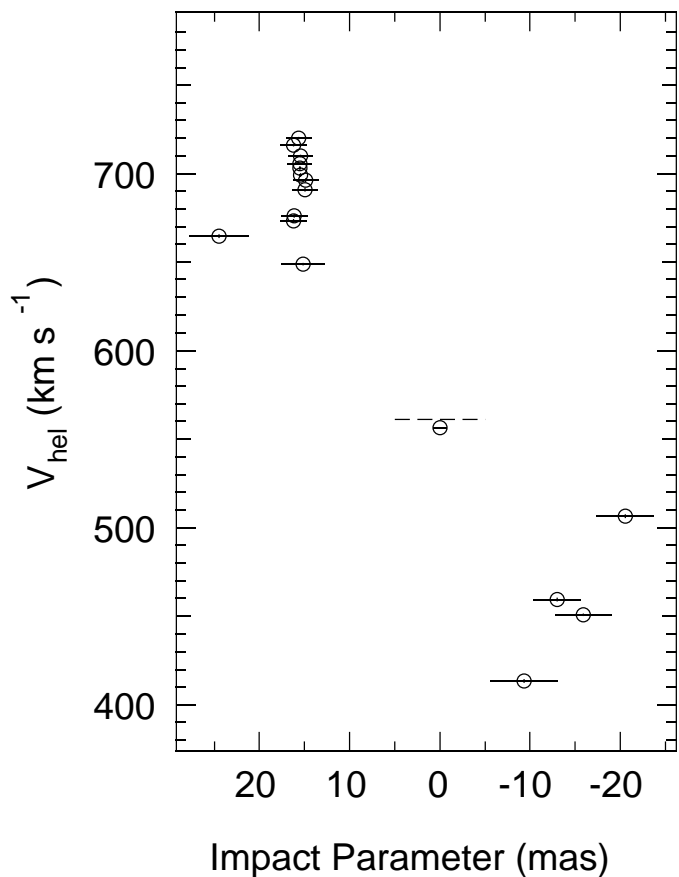}{4.2in}{0}{100}{100}{-320}{-210}
\caption{The velocities of the mapped maser features in NGC 4945 as a
function of impact parameter. The positions are measured relative to
the feature that is closest in velocity to the systemic velocity
(dashed line). The error bars on position are $3\sigma$. From Greenhill
\etal 1997.}  
\label{fig8}
\end{figure}

\subsection{NGC 3079}
NGC 3079 is a relatively nearby ($D=16$ Mpc), nearly edge-on spiral
galaxy. It has a fairly high far-infrared luminosity ($L_{\rm
IR}\approx 3\times 10^{10}\lsol$), a LINER-type spectrum (Heckman
1980) and a spectacular nuclear outflow (see Cecil \etal 2001, and
references therein), with prominent radio lobes (Duric \& Sequist
1988) and loops of H$\alpha$ emission (Ford \etal 1986) extending a
kpc or more away from the nucleus along the minor axis. NGC 3079 also
has a prominent ring of molecular material, seen in CO emission,
extending from approximately 250 to 750 pc, and a very compact,
luminous CO peak centered on the nucleus (Sofue \etal 2001, and
references therein). 

NGC 3079 possesses one of the most luminous \htwoo\ megamasers known
(Henkel \etal 1984a,b; Haschick \& Baan 1985). The maser emission is
almost entirely concentrated to velocities blueward of the systemic
velocity, and it is strongly variable in flux density and line width,
the flux density having varied by about a factor of eight over ten
years, while the linewidths have fluctuated by up to a factor of two
(Baan \& Haschick 1996).

The maser emission from NGC 3079 was observed using the VLBA by
Trotter \etal (1998). The observed distribution of maser emission on
the sky, color-coded according to velocity, is shown in Figure 9. The
two dashed lines show the major axis of the molecular disk discussed
above (the nearly vertical line) and the axis of the nuclear jet. The
emission arises in compact clumps, with sizes $\lessapprox 0.02$ pc;
most of the emission arises in a region less than 0.2 pc in diameter,
spread over a velocity range of about 130 \kms. The size of the
circles representing the maser emission are proportional to the
logarithm of the flux density, and they are labeled with the velocity
of the approximate flux density peak. The inset in Figure 9 (from
Trotter et al.) shows an expanded version of the concentrated
emission; this region includes the maser peak, at $v_{\rm
LSR}=957\kms$. Also sketched in Figure 9 are the positions and
approximate sizes of 22 GHz continuum features.

\begin{figure}
\plotfiddle{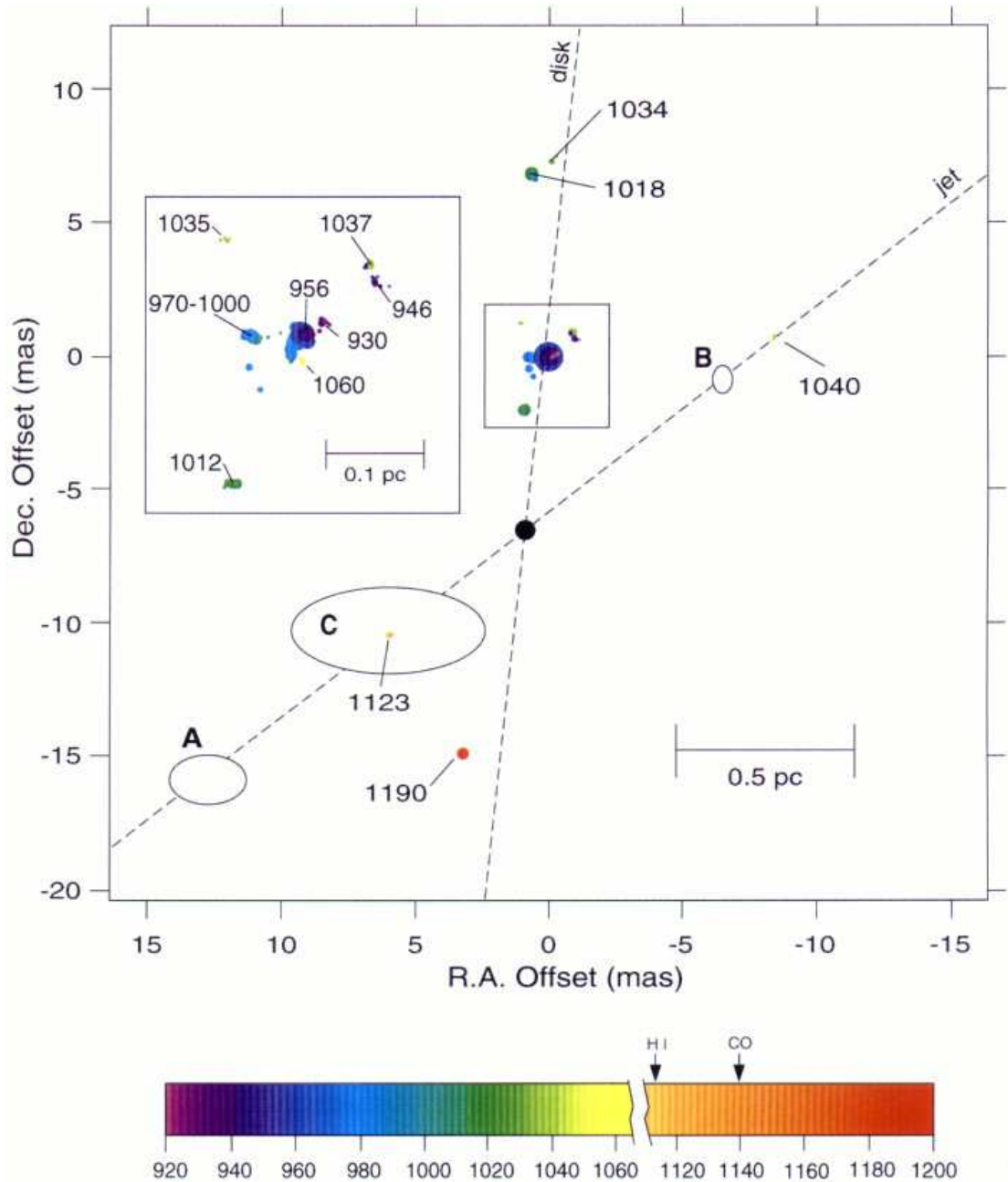}{6.0in}{0}{75}{75}{-225}{-20}
\caption{The spatial and velocity distribution of the maser emission
in NGC 3079, as observed with the VLBA, from Trotter \etal (1998). The
spots marking the maser features are sized according to the logarithm
of the flux density and color-coded according to velocity; the labels
show the velocity of the approximate flux density peak. Also sketched
in are the approximate positions and sizes of 22 GHz radio continuum
features, and the axes of the large-scale molecular disk and the
nuclear radio jet. }
\label{fig9}
\end{figure}

It is evident from Figure 9 that the maser emission from NGC 3079
differs markedly from the distribution seen in NGC 4258. The maser
spots are approximately aligned with the larger-scale molecular disk,
and their measured Doppler velocities are consistent with rotation in
the same sense as the molecular disk. However, there is evidently a
large nonrotational component to the velocity field: as noted above,
the spread in velocities within the compact grouping of maser spots
covers the entire range of blueshifted emission. This may indicate
large turbulent velocities in the disk. The 22~GHz continuum emission
is dominated by an unresolved ($\lessapprox 0.1$ pc) source (labeled B
in Figure 9) that lies approximately 0.5 pc to the west of the maser
peak. There is no maser emission associated with this (or any other)
continuum peak; this argues against the suggestion that the unusually
luminous maser emission in NGC 3079 is the result of beamed,
unsaturated amplification of compact continuum emission by foreground
molecular gas.

A sketch showing the model of the emission suggested by Trotter \etal
(1998) is shown in Figure 10; this must be regarded as highly
speculative (see below). As with NGC 4945, the emission has been
interpreted within the framework established by NGC 4258, \ie as a
rotating disk, seen close to edge-on. The disk thickness is
unknown. Trotter \etal argued that the maser emission could not be
powered by X-ray irradiation, as the observed X-ray luminosity of NGC
3079 was only $L_x\sim 10^{40}\>\ergpsec$ in soft X-rays, and there
was no indication of a nuclear hard X-ray source. However, {\it
BeppoSax} observations (Iyomoto \etal 2001) show that, as in the
case of NGC 4945, there is a luminous ($L_x\sim
10^{42}-10^{43}\>\ergpsec$) hard X-ray source buried beneath
Compton-thick absorption, thus eliminating this argument. The overall
distribution of velocities is consistent with the presence of a
binding mass $M\sim 10^6\msol$ within 0.5 pc. If this is indicative of
the central mass, the hard X-ray luminosity indicates that the AGN
contributes a large fraction of the far-infrared luminosity of the
nucleus of NGC 3079, and it is radiating at $1-10\%$ of the Eddington
limit.

Sawada-Satoh \etal (2000) argue that this interpretation is not
correct, as they identify the bright continuum source (labeled ``B''
in Figure 9) as the location of the nucleus. This identification is
based on the apparent fixed position of component B with respect to
the strongest maser feature based on several years of observations,
unlike components A and C, which they argue are probably jet
features. Sawada-Satoh \etal suggest that the maser emission arises in
a geometrically thick structure whose rotation axis passes through
component B; this axis is inclined by about $110^\circ$ to the
large-scale rotation axis of the galaxy.

\begin{figure}
\plotfiddle{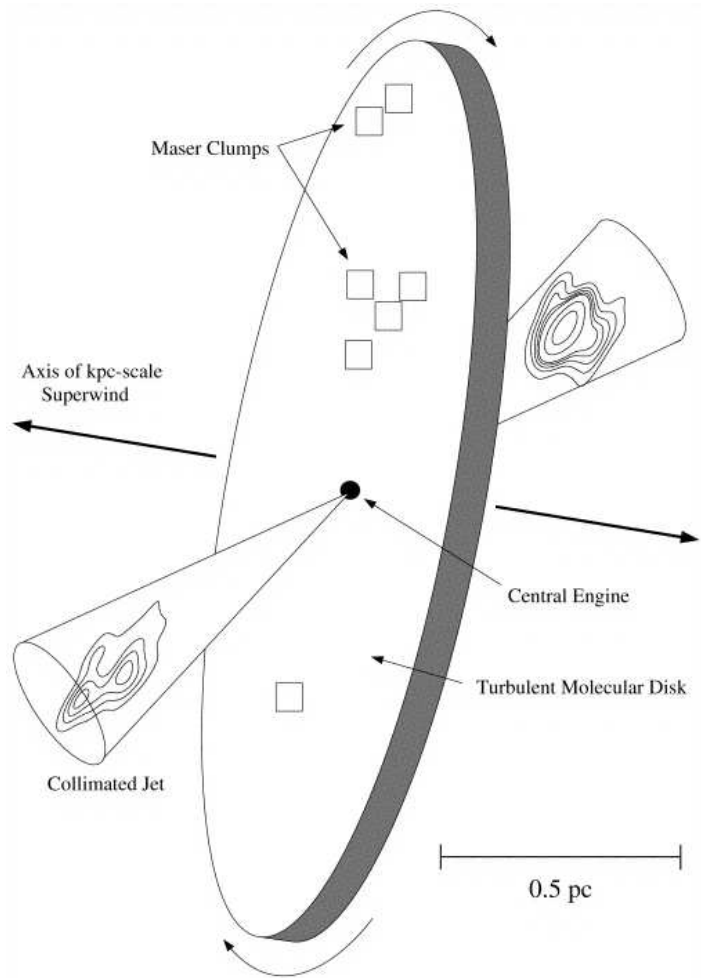}{5.0in}{0}{65}{65}{-195}{-70}
\caption{Model of the central pc of NGC 3079, based on VLBA
observations (from Trotter \etal 1998). The maser spots are
distributed in a disk that is close to edge-on; the disk center is
assumed to be near where the major axis of the maser emission and the
jet axis intersect.}
\label{fig10}
\end{figure}

\subsection{Circinus}
The Circinus galaxy is a nearby ($D=4$ Mpc) Seyfert 2 galaxy in the
Southern hemisphere, and was one of the first megamaser sources to be
discovered (Gardner \& Whiteoak 1982). It bears interesting
similarities to several of the galaxies discussed above, including the
presence of a luminous ($L_x\sim 10^{42}\>\ergpsec$) hard X-ray source
hidden behind Compton-thick obscuration (Matt \etal 1999), a nuclear
outflow along the minor axis (seen in bipolar radio lobes: Elmouttie
\etal 1998), a broad (opening angle $\sim 90^\circ$) ionization cone
(Marconi \etal 1994; Veilleux \& Bland-Hawthorn 1997) and a nuclear
($\sim$ several hundred pc) molecular ring seen in CO emission (Curran
\etal 1998). The maser spectrum shows emission out to approximately
$\pm 200\kms$ with respect to the systemic velocity; the red-shifted
emission is stronger than the blue, and both are substantially
stronger than the systemic (Nakai \etal 1995; Braatz, Wilson, \&
Henkel 1996; Greenhill \etal 2001). The maser emission is highly
variable, and the spectrum alters substantially on timescales of order
one month (Whiteoak \& Gardner 1986). Variability has been seen on
timescales as short as a few minutes (Greenhill \etal 1997b), much more
rapidly than any other source. This rapid variability is almost
certainly the result of interstellar scintillation; Circinus lies at
low Galactic latitude, $b=-3.8^\circ$.

\begin{figure}
\plotfiddle{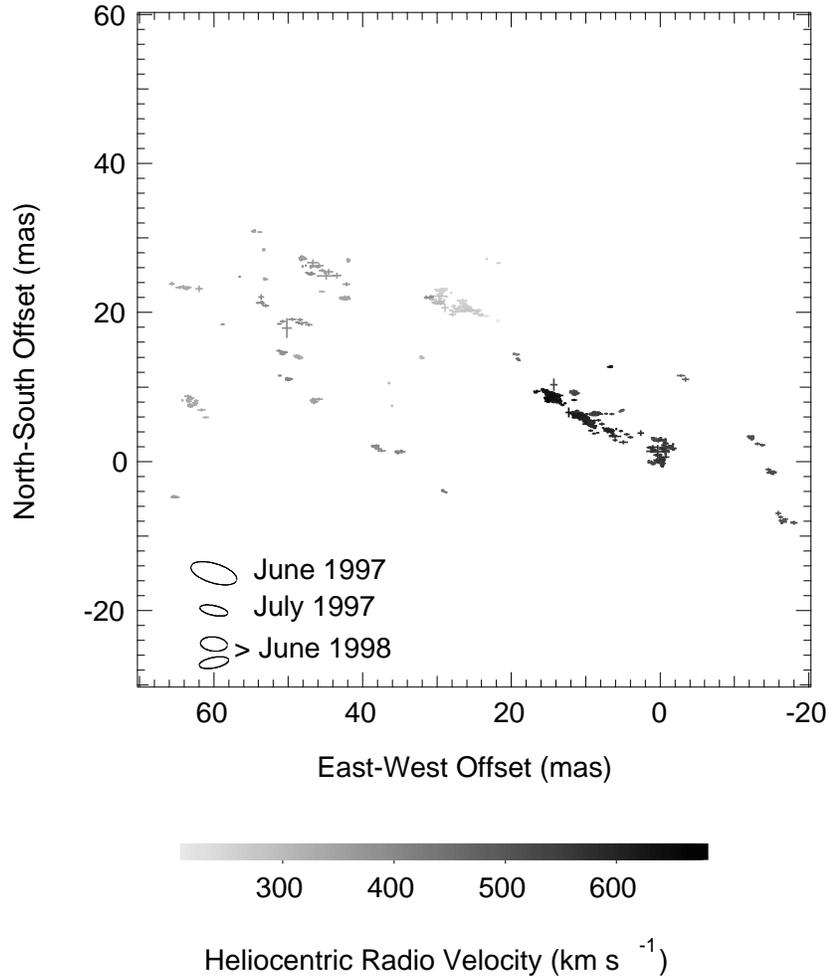}{4.8in}{0}{65}{65}{-200}{-100}
\caption{The spatial distribution of \htwoo\ maser emission from the
Circinus galaxy. The observations were actually taken at three
separate epochs; the synthesized beam sizes are plotted at lower
left. The velocity of the emission is indicated by the gray scale. At
the distance of Circinus, $1\ {\rm mas} = 1.9\times 10^{-2}$ pc. The
systemic velocity $V_{\rm sys}\simeq 440\kms$. From Greenhill \etal
(2001).}  
\label{fig11}
\end{figure}

\begin{figure}
\plotfiddle{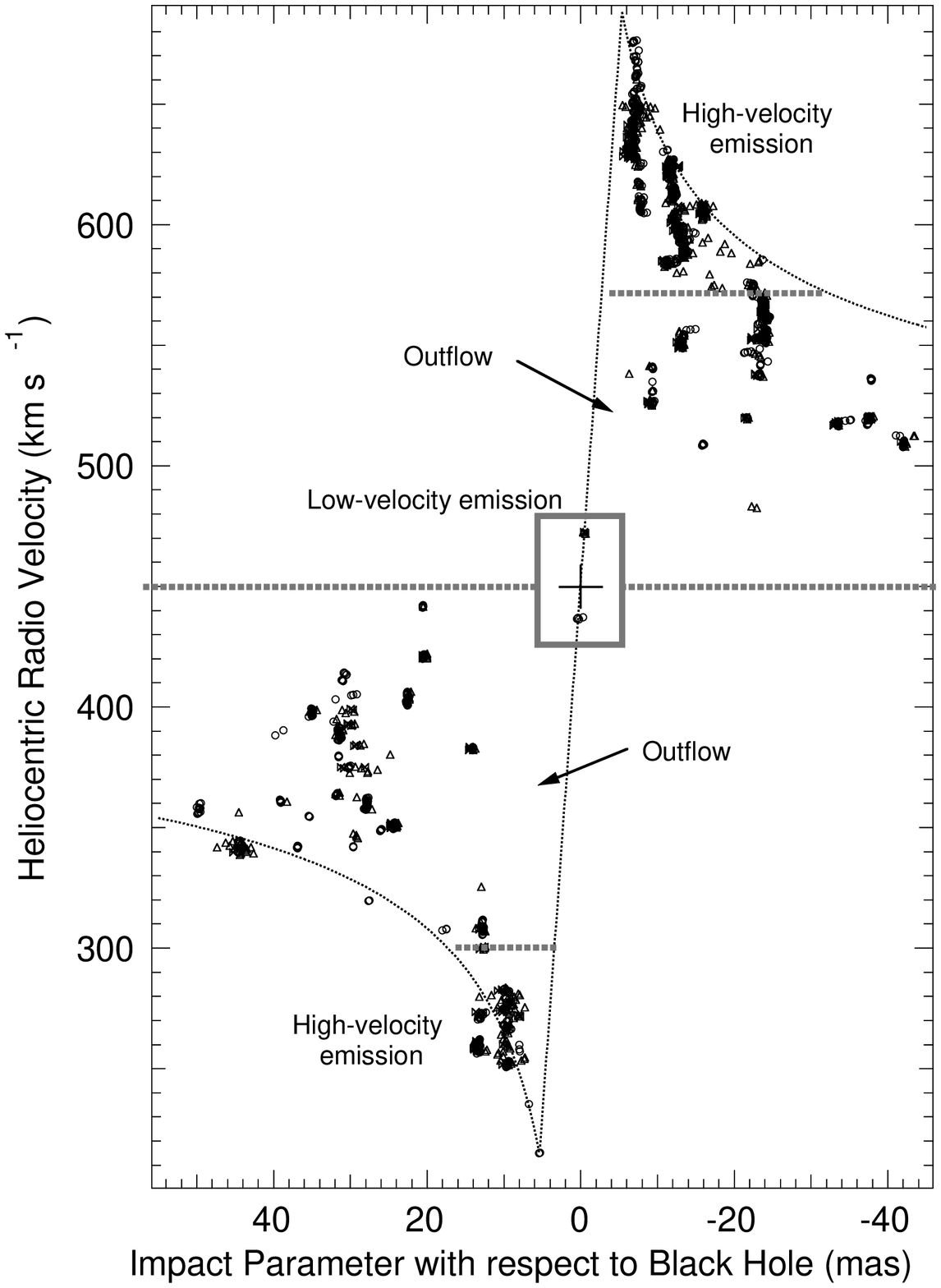}{5.3in}{0}{70}{70}{-210}{-50}
\caption{Position-velocity diagram for the Circinus maser
emission. Labels indicate features of the model discussed in the
text. The rotation curve declines with impact parameter $b$
approximately as $b^{-0.4\pm 0.05}$. From Greenhill (2001b).} 
\label{fig12}
\end{figure}

VLBI observations of the \htwoo\ maser emission from Circinus have
been reported by Greenhill (2001b) and Greenhill \etal (2001). The
emission was observed using the Australia Telescope Long Baseline
Array in 1997 and 1998. The distribution on the sky (with gray scale
indicating velocity) is shown in Figure 11. Although this appears
somewhat confusing, when combined with the velocity information
(Figure 12) a coherent picture emerges. At each observing epoch, the
maser emission consists of three components: a thin, shallow $\cal{S}$
shape on the sky (similar to NGC 4945), comprised of highly redshifted
and blueshifted emission; emission at close to the systemic velocity,
which lies between the two arcs of high-velocity emission; and
modestly Doppler-shifted (up to $\sim 100\kms$) emission whose spatial
distribution is correlated with the sign of the velocity shift:
blueshifted emission on the southeast side, redshifted on the
northwest side. 

The high-velocity and near-systemic velocity emission can be fitted by
a thin (thickness $< 2\;{\rm mas} = 0.04$ pc), warped disk, viewed
edge-on, with an outer radius of approximately 0.4 pc (20 mas) and an
inner edge at $\sim 0.1$ pc. The orbital velocity (assuming circular
motion) at the inner edge is 237 \kms, giving an enclosed mass
$M\simeq 1.3\times 10^6\msol$, and a minimum mass density $\rho \sim
3\times 10^8\msol\pcubpc$. With this central mass, the observed X-ray
luminosity implies a fractional Eddington luminosity of order
$10\%$. The high-velocity emission on the redshifted side can be
fitted with a peak velocity that declines with impact parameter $b$
approximately as $b^{-0.4\pm 0.05}$. The fact that the rotation curve
is shallower than Keplerian could be due to the mass of the disk; this
would imply a disk mass $\sim 10^5\msol$ between 0.1 and 0.4
pc. Alternatively, this could be due to the contribution of a central
stellar cluster. Lending additional support to this model is the fact
that a single position-velocity gradient connects the highest velocity
red- and blue-shifted emission, and the near-systemic velocity
emission; this is also seen in NGC 4258, and is expected if all of the
near-systemic emission arises at the inner radius of the disk
(although this emission is relatively much weaker than in NGC 4258).

What does the intermediate-velocity gas represent? Greenhill (2001b)
and Greenhill \etal (2001) interpret this as a fairly broad-angle
outflow (see Figure 13). The systematic redshifting and blueshifting
with position on the sky indicate that the outflow is tipped with
respect to our line of sight. The high-velocity emission ceases where
it would be intersected by the outflow, which may indicate that the
disk is broken up by the interaction. Interestingly, all of the
outflow maser emission arises in regions where the view of the central
source would not be shadowed by the disk, which suggests that the
maser emission is due to irradiation of the outflow by hard X-rays
from the central source. Furthermore, the edges of the shadowed zone
coincide very well with the boundaries of the kiloparsec-scale
ionization cone seen to the west of the nucleus. However, truncation
of the disk also implies that the central black hole will be starved
by exhaustion of the disk mass in no more than $\sim 10^7$ years
(Greenhill 2001b).

\begin{figure}
\plotfiddle{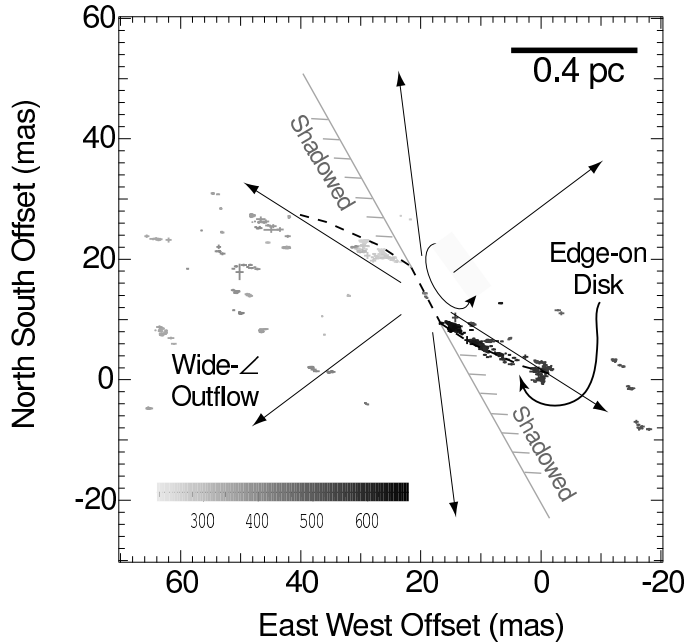}{3.0in}{0}{70}{70}{-210}{-180}
\caption{Model of the warped accretion disk and outflow structure in
the Circinus galaxy, superimposed on the spatial distribution of the
emission. From Greenhill (2001b).}
\label{fig13}
\end{figure}

\subsection{IC 2560}
IC 2560 is a southern ($\delta\approx -33^\circ$) barred spiral in the
Antlia cluster, at a distance of 26 Mpc. It is classified as a Seyfert
2 on the basis of optical spectra (Fairall 1986). \htwoo\ maser
emission was first detected by Braatz \etal (1996); the emission
velocity was close to the systemic velocity of the galaxy.

The maser emission was reobserved by Ishihara \etal (2001), using both
single-dish telescopes (the NRO 45m and the Parkes 64m dishes) and the
VLBA. Multiple epochs were obtained, with observations spanning the
period 1996--2000. Most of the flux is contained in the emission
around the systemic velocity, which consists of a number of narrow
individual features blended together over a band $\Delta V\approx
60\kms$ wide. (This was the only emission detected by Braatz et al.)
The peak flux density is variable and has reached a maximum of about
twice the value at the time of the Braatz \etal observation. Most
interestingly, however, is the detection of additional, high-velocity
emission, extending to approximately $400\kms$ away from $V_{\rm sys}$
to both the red and the blue (see Figure 14). The redshifted emission
is only slightly stronger than the blueshifted; the brightest
high-velocity emission features are about $12-14\%$ of the peak flux
density. 

\begin{figure}
\plotfiddle{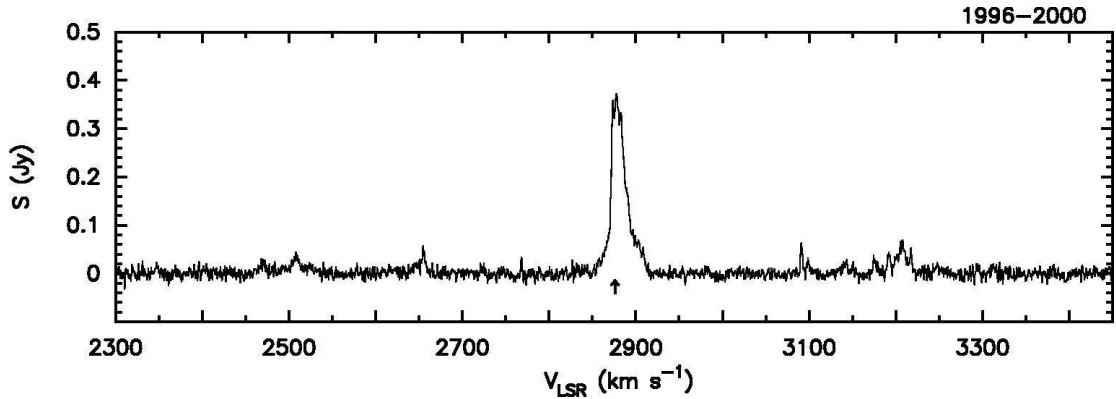}{2.0in}{0}{75}{75}{-225}{-10}
\caption{Spectrum of the \htwoo\ maser emission from IC 2560 obtained
with the NRO 45m telescope, averaged over all epochs. The arrow marks
the systemic velocity of the galaxy. From Ishihara \etal (2001).}
\label{fig14}
\end{figure}

The spectrum of IC 2560 bears remarkable similarity to that of NGC
4258 (compare Figure 14 with the top panel of Figure 4), in that there
are weaker, high-velocity features spread over a substantial range to
either side of the systemic emission, although the velocities are not
as high in IC 2560, and the high-velocity emission is more
symmetrically distributed than in NGC 4258. This is the only megamaser
observed to date other than NGC 4258 in which the systemic velocity
emission is much stronger than the high-velocity emission. This is not
the only similarity, however. Ishihara \etal claim that over the
course of the monitoring period (4.4 years using the NRO 45m and 1.1
year with the Parkes 64m), the systemic velocity features showed a
secular drift to redward, at a rate $\dot v \simeq 2.6\kms\pyr$. In
NGC 4258 this drift (at $\dot v\simeq 9\kms\pyr$) is interpreted as
centripetal acceleration, as discussed in \S 3. A similar velocity
drift has been seen in NGC 2639 (Wilson \etal 1995), with $\dot
v\simeq 6.6\kms\pyr$. One high-velocity feature was also monitored
over a period of about a year and showed no secular acceleration, with
an upper limit $\dot v \lessapprox 0.5\kms\pyr$.

Given the spectral similarities, one might expect imaging of the maser
emission to be equally intriguing. Unfortunately, only the systemic
velocity emission has been observed with VLBI so far, as the
high-velocity emission had not been detected at the time of the VLBA
observation. Furthermore, the high-velocity emission is faint, with
flux densities $\sim 0.05 0.1$ Jy, so that imaging this emission will
be very difficult. Nevertheless, the results are indeed
interesting. Figure 15 (from Ishihara \etal 2001) shows the spatial
distribution of the emission and position-velocity cuts in both right
ascension and declination. The emission is confined to a region
approximately 0.1 mas by 0.2 mas in RA and Dec ($0.01\times 0.02$ pc
at 26 Mpc); note that the position errors in declination are much
larger than those in RA. This is similar to the spatial extent of the
systemic velocity emission in NGC 4258. Furthermore, the
position-velocity diagram in RA shows a systematic, linear trend, as
in NGC 4258, with a magnitude $dv/db\simeq 0.85\pm 0.27\kms\;\mu{\rm
as}^{-1}$. (No trend is seen in the p-v diagram in Dec, which may be
due to the large position errors or to the orientation.) There is also
an unresolved continuum source detected at 22 GHz, with a very high
brightness temperature ($T_b \gtrapprox 3\times 10^{11}$ K), which is
coincident with the grouping of maser emission within the errors.

\begin{figure}
\plotfiddle{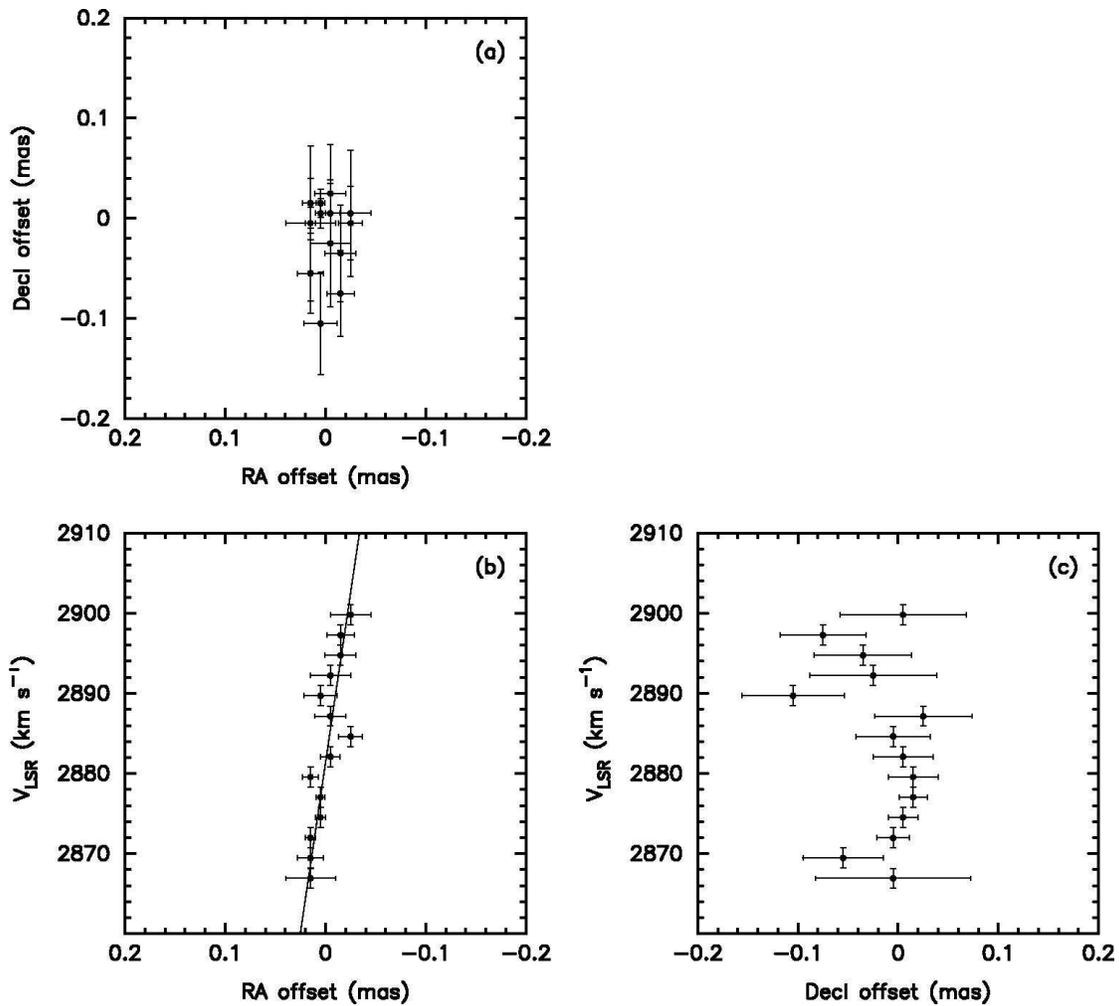}{5.3in}{0}{75}{75}{-225}{30}
\caption{Spatial and velocity distribution of the systemic-velocity
maser emission from IC 2560. (a) shows the position of the maser
emission on the sky. (b) and (c) show position-velocity diagrams for
cuts in right ascension and declination, respectively. The RA p-v cut
shows clear evidence for a systematic, linear gradient of velocity
with impact parameter. From Ishihara \etal (2001).}
\label{fig15}
\end{figure}

The maser emission in IC 2560 is certainly suggestive, although the
rotation curve has not yet been obtained. When placed in the framework
of an edge-on, Keplerian disk, the observations imply a disk with
inner and outer radii of 0.07 and 0.26 pc, respectively, in orbit
around a central mass of $M=2.8\times 10^6\msol$, about an order of
magnitude smaller than that for NGC 4258 or NGC 1068, but comparable
to the central masses inferred for NGC 4945, Circinus, and NGC
3079\footnote{With this mass, the inner radius of the maser disk is at
$R\simeq 2.6\times 10^5$ Schwarzschild radii, while the outer edge
lies at nearly $10^6 R_s$.}. It must be admitted that the distribution
of systemic emission on the sky does not appear disk-like. However, it
must also be remembered that IC 2560 is about 3.5 times further away
than NGC 4258, so that (along with the smaller inner radius) the
angular scale of the emission is smaller by about a factor of
six. Furthermore, the emission is several times weaker than in NGC
4258, so that the positional error bars are quite large, especially in
declination, where they are $50-60$ $\mu$as.

With a maximum rotation velocity of $418\kms$, the velocity drift of
the systemic velocity features places them at $0.07$ pc, which
Ishihara \etal identify with the inner radius of the disk. If this is
correct, the lower limit to the central mass density is
$\rho\gtrapprox 2.1\times 10^9\msol\pcubpc$, second only to NGC 4258
among the maser disks observed to date. As in NGC 4258, this location
for the systemic velocity features is also supported by the gradient
seen in the position-velocity diagram: for a given enclosed mass $M$,
the radius implied for a measured $dv/db$ (where $v$ is the projected
velocity and $b$ is the impact parameter), for a Keplerian rotation
curve, is \be R=\left[{(GM)^{1/2}\over dv/db}\right]^{2/3}= 4.66\times
10^{-3} {M_6^{1/3} d^{2/3}_{\rm Mpc}\over (dv/db)^{2/3}}\ \ {\rm pc}
\ee where $M_6$ is the mass of the central black hole in units of
$10^6\msol$, $d_{\rm Mpc}$ is the distance to the host galaxy in Mpc,
and the velocity gradient $dv/db$ is in $\kms\;\mu{\rm as}^{-1}$. For
IC 2560, equation (16) yields $R=0.064\pm 0.02$ pc, in excellent
agreement with the value estimated from $\dot v$. This agreement also
implies that the disk major axis is aligned within $\sim 20^\circ$ of
east-west. The velocity extent of the systemic emission implies (in
the Keplerian disk framework) a very similar beaming angle to NGC
4258, about $9^\circ$. 

There is one additional similarity between IC 2560 and NGC
1068. Ishihara \etal (2001) analyzed the X-ray emission from IC 2560,
which was observed by {\it ASCA} in 1996 for approximately 42 ksec
with the GIS and SIS spectrometers. Although the S/N is poor, Ishihara
\etal find evidence for a heavily-absorbed ($N_H\sim 3\times
10^{23}\psqcm$) hard X-ray source, with a 2--10 keV X-ray luminosity
$L_x\sim 10^{41}\>\ergpsec$. It could be substantially lower than this;
a much higher S/N spectrum is needed. Assuming that this represents
$10\%$ of the bolometric luminosity, the fractional Eddington
luminosity is of order $10^{-3}$, which classes IC 2560 with NGC 4258
as having very low $L/L_{\rm Edd}$, as opposed to the other possible
maser disk systems with $L/L_{\rm Edd}\sim 0.1$ or greater. If we make
the assumption that the maser emission is distributed in a thin,
Keplerian disk that is irradiated by the central X-ray source, as in
NGC 4258, and apply the theory of NMC to identify the outer edge of
the disk with the molecular-to-atomic phase transition (equation 4),
we obtain a mass accretion rate $\dot M/\alpha\sim 8\times
10^{-4}\msol\pyr$, and with the above estimate of the luminosity, a
radiative efficiency $\epsilon\sim 0.02/\alpha$, indicating that the
luminosity of IC 2560 is low simply because, as in NGC 4258, the mass
accretion rate is low. Further observations of this system with VLBI
are eagerly awaited. If the geometry is confirmed to be similar
to the NGC 4258 disk, it will be possible to obtain another purely
geometric measurement of the distance to the maser disk, but for a
galaxy which is nearly four times further away than NGC 4258, which
would be very important for determination of $H_o$ (as the
distance determination for NGC 4258 already has been, since it allows
one to simply skip over all of the argument about the distance modulus
to the LMC: Newman \etal 2001). 

Of the sources for which available VLBI imaging data provide evidence
for maser disks, only the emission from IC 2560 appears to bear a marked
similarity to the NGC 4258 disk (although until the recently
discovered high-velocity emission from IC 2560 is imaged, the nature
of the rotation curve will remain unknown). Is this surprising? Not
necessarily, at least in one aspect. 

Consider a galaxy containing a supermassive black hole in its
nucleus. The black hole dominates the potential within a radius
\be
R_{BH}\approx 2 {M_7\over \sigma_{150}^2}\;{\rm pc}
\ee
where the velocity dispersion characterizing the depth of the galactic
potential well (not including the contribution from the black hole) is
$\sigma=150\sigma_{150}\kms$. Using the X-ray powered maser model of
NMC and NM95, the critical radius for molecule formation can be
written in terms of the fractional Eddington luminosity (\cf equation
4) as
\be
R_{cr}= 1.5{(L/L_{\rm Edd})^{0.38} M_7\over (\alpha\epsilon_{0.1})^{0.81}
\mu^{0.38}}\;{\rm pc}
\ee
where the radiative efficiency $\epsilon =0.1\epsilon_{0.1}$.
In order for a maser disk to exhibit a Keplerian rotation
curve, it must satisfy $R_{cr}\ll R_{BH}$, or
\be
{L\over L_{Edd}}\ll 0.5 {(\alpha\epsilon_{0.1})^{2.1}(\mu/0.25)\over
\sigma_{150}^{5.3}}.
\ee
Hence only AGN with very sub-Eddington luminosities will have maser
disks in the Keplerian regime. Since, as noted above, IC 
2560 appears
to fulfill this requirement, with $L/L_{\rm Edd}\sim 10^{-3}$, it will
be very interesting to see what VLBI reveals about the structure of
the high-velocity emission.

\section{Imaging 3: The Non-Disk Sources}
Nearly all of the \htwoo\ megamasers detected so far are in
disk-dominated galaxies. There are two exceptions to this, which have
interesting similarities to one another and also to a third which has
yet to be fully imaged. The three galaxies (the first two of which are
listed in Table 1) are NGC 1052 (an elliptical galaxy with a LINER
nucleus), IRAS 22265, also known as TXFS 2226-184, originally
classified as an elliptical or S0 but now definitely known to be a
disk galaxy (Falcke \etal 2000a), and Mrk 348, classed as an S0. A
fourth object, IRAS F01063-8034, which is definitely a disk galaxy,
has a recently discovered maser (Greenhill \etal 2002) which also
appears to be similar, and will be discussed briefly below.

These four objects stand out from the other megamasers on the basis
of their line profiles. As noted in \S 1, the typical line profile for
megamasers consists of a number of narrow components, with widths of a
few \kms, spread over a much broader velocity range, $\Delta V\sim
10^2-10^3\kms$. The four galaxies discussed in this section have
completely different profiles, consisting of single, broad ($\Delta
V\approx 90-130\kms$, except for IRAS F01063, for which $\Delta
V\approx 40-50\kms$), more or less Gaussian lines. In the cases of NGC
1052 and Mrk 348, the maser emission is offset by approximately
$150\kms$ with respect to the systemic velocity of the galaxy; no such
offset has been reported for the IRAS 22265 emission and the IRAS
F01063 emission is at the same velocity as the galaxy.

The spectral differences alone suggest that these masers differ in a
fundamental way from the other megamaser sources. Imaging with VLBI
appears to confirm this, at least for the first two mentioned.
Claussen \etal (1998) presented VLBI observations of NGC 1052 in both
22 GHz continuum and the $6_{16}-5_{23}$ water line. The continuum
data show an elongated structure composed of at least seven
components, with a gap near the center. The maser emission occurs in
two moderately distinct groups, separated by about 0.02 pc, with a
total extent of about 0.06 pc. Interestingly, the maser emission
appears to lie along the axis of the ``jet'' seen in the continuum
emission. There is also a clear velocity gradient in the maser
emission, also in the direction parallel to the jet axis. The masers
may be excited by shocks driven by the interaction of the jet with
molecular clouds, as may be seen in the ``jet masers'' of NGC 1068
(Gallimore \etal 2001). Alternatively, given the superimposition of
the masers on the continuum emission, the masers may arise in
foreground molecular clouds that are amplifying the continuum emission
(although, given the very small angular and spatial scales involved,
the absence of maser emission from the vicinity of the other continuum
components is a little surprising in this interpretation). The latter
scenario can be tested by searching for correlated variation of the
continuum and maser emission (Claussen et al.), since for this model to
work the maser emission must be unsaturated. The velocity centroid of
the maser emission shifted by $45\kms$ between two observations spaced
five months apart (Braatz \etal 1996), which argues in favor of the
jet-excited model.

IRAS 22265 has been imaged with VLBI but the results have not yet been
published (Greenhill, Moran, \& Henkel 2002, in preparation). In Moran
\etal (1999), from which Table 1 has been adapted, the distribution of
maser emission is described as ``messy'', with a spatial extent of
about 2.4 pc (note that their Table 3 contains a minor typographical
error: the spatial extent of the emission in NGC 1052 is given as 0.6
pc rather than 0.06 pc). Greenhill \etal (2002) note that the
emission from this source has been resolved into at least four
components, spread over roughly a parsec, with individual line widths
of tens of kilometers per second.

\begin{figure}
\plotfiddle{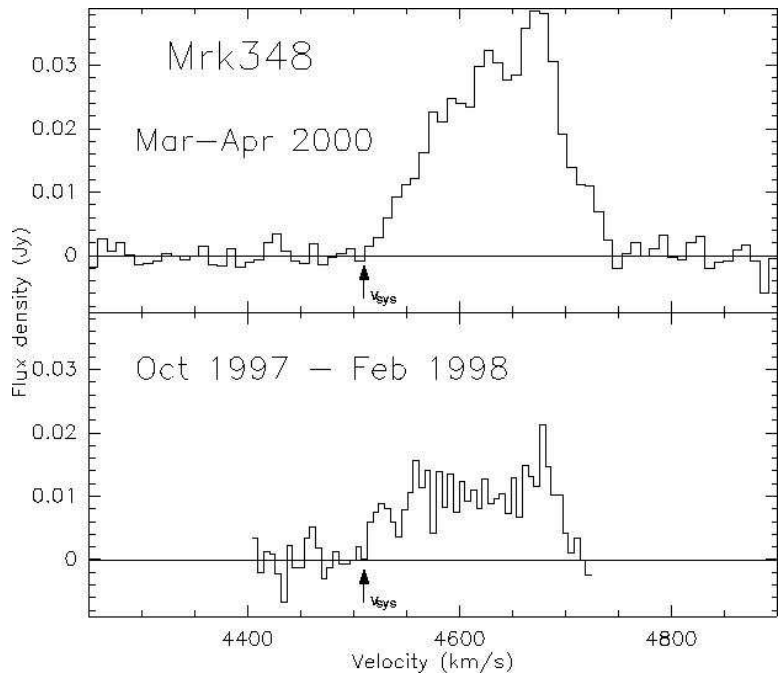}{3.4in}{0}{52}{52}{-170}{-5}
\caption{Spectrum of the maser emission from Mrk 348, at two different
epochs. The maser clearly rose in intensity by a factor of three
between late 1997 and early 2000. Also note the redshift of the
emission with respect to the systemic velocity. From Falcke \etal
(2000b).}
\label{fig16}
\end{figure}

The third source with a single, broad maser line, Mrk 348, is a
Seyfert 2 galaxy, with broad emission lines in polarized light. It is
unusual among Seyfert galaxies in having a very bright and variable
radio nucleus. Among the megamaser sources, in addition to its line
profile, it is notable for having flared by about a factor of three in
its 22 GHz emission (Falcke \etal 2000b), sometime between the end of
1997 and early 2000 (see Figure 16). It was observed but not detected
in the Braatz \etal survey; analysis of archival observations by
Falcke \etal show that it would have been seen only at the $3\sigma$
level at the time of the Braatz \etal observations. Furthermore,
Falcke \etal present evidence that the radio continuum emission in Mrk
348 also rose between late 1997 and early 2000 by about the same
factor as the maser emission. (Between early 1997 and late 1998, the
flux density at 15 GHz was observed to increase by nearly a factor of
5 by Ulvestad \etal 1999.) If the variations of continuum and line are
in fact correlated, this would indicate that the maser emission in Mrk
348 is unsaturated. The upper limit to the response time lag between
the radio flare and the maser flare (about two years) sets an upper
limit to the distance of the masers from the nucleus of about 0.6
pc. A recent VLBI observation with MERLIN (Xanthopoulos \& Richards
2001) of the red half of the line observed by Falcke \etal confirms
that maser emission arises within $\sim 0.8$ pc of one of the
continuum peaks; the peak flux density is about three times higher
than when observed by Falcke \etal (2000b), so that the maser flux
density has increased by about an order of magnitude since late 1997.

The fourth source, IRAS F01063-8034, is a moderately distant (57 Mpc)
southern edge-on disk galaxy, classified as Sa, which shows no real
sign in the optical, infrared, or radio of possessing an active
nucleus. Greenhill \etal (2002) discovered luminous ($L_{\rm H_2O}\sim
450\lsol$ maser emission from this galaxy. The maser emission consists
of a single Gaussian line, with a FWHM $\Delta V\approx 54\kms$. As in
NGC 1052, the maser emission exhibited a shift in velocity, with the
centroid jumping by 38 \kms\ within thirteen days. As with NGC 1052,
such velocity shifts are difficult to understand except in the context
of jet-excited models. In the case of IRAS F01063, this offers
evidence for the presence of a previously unsuspected AGN in this
galaxy.

The similarities among these four sources suggest that a common
physical process is at work. The broad, smooth emission profiles may
indicate that the maser action in all four of these galaxies is
unsaturated, which would agree with the superimposition of the maser
emission on the radio continuum source in NGC 1052, and with a
correlated continuum/maser flare in Mrk 348. The spatial distribution
of the emission, which is less cohesive than seen in most of the other
megamasers imaged to date, may indicate a different origin than X-ray
pumping, such as shocks, although this is not yet clear: both NGC 1052
and Mrk 348 are known to possess luminous ($L_x\sim 10^{42}-10^{43}\>
\ergpsec$), heavily obscured ($N_H\sim 10^{23}\psqcm$), nuclear hard
X-ray sources (Weaver \etal 1999; Warwick \etal 1989). Certainly the
velocity shifts seen in both NGC 1052 and IRAS F01063-8034 argue for
a shock origin for the masers, perhaps due to jet/gas interactions.

\section{The Future}
Future, deeper surveys for \htwoo\ megamasers can be expected to turn
up many more sources in Seyfert 2 and LINER galaxies, given the known
detection fractions (Braatz \etal 1996, 1997). Unfortunately, with
existing instruments very few of these additional sources will be
bright enough to image with VLBI; with the VLBA it is difficult
(although not impossible, as discussed earlier) to observe sources
that are fainter than about 0.5 Jy. The advent of the Atacama Large
Millimeter Array (ALMA) will offer a very large increase in
sensitivity over the VLA. However, ALMA will not be able to observe
the 22 GHz line, as the lower frequency limit will be 30 GHz, and, in
any case, would not have the spatial resolution at this frequency ($<
1$ mas) required to image the nuclear maser emission in distant
galaxies. The proposed Square Kilometer Array (SKA) will have a
huge collecting area, and consequently enormous sensitivity at 22 GHz;
hence SKA will be able to detect very faint, distant megamaser
sources. However, it is not certain that SKA will be designed with the
capability of performing high-resolution observations, in which case
it will be superb for maser surveys but will not be able to do the
follow-up imaging work. The Green Bank Telescope (GBT) will of course
not have the necessary spatial resolution for imaging, but will be a
superb instrument for monitoring studies.

However, imaging in the 22 GHz line is not the only method by which
our understanding of \htwoo\ megamasers can be improved. The only
transition that has been observed so far from these sources is the
$6_{16}-5_{23}$ line at 22 GHz. However, a number of other water lines
are predicted to mase, and several other masing transitions have been
detected in Galactic sources: the $3_{13}-2_{20}$ line at 183 GHz
(Waters \etal 1980; Cernicharo \etal 1990); the $10_{29}-9_{36}$ line
at 321 GHz (Menten, Melnick, \& Phillips 1990); and the
$5_{15}-4_{22}$ line at 325 GHz (Menten \etal 1990). In all of these
sources the 22 GHz line is seen as well, and it is very likely that
the maser radiation is arising in the same gas (\eg the velocity range
of emission is the same for all the transitions, except where
detection is limited by low signal-to-noise; the line ratios seen in
these sources, all of which are single-dish observations, are in
reasonable agreement with model predictions; see Menten \etal
1990). All of these transitions are substantially weaker (both
observationally and theoretically) than the 22 GHz line. However, the
very large increase in sensitivity provided by ALMA will make it
possible to observe the 321 and 325 GHz (and, under very favorable
atmospheric conditions, the 183 GHz transition) from the ground.  At
325 GHz, ALMA will have spatial resolution of 20~mas (assuming the
planned 10 km baseline).

As discussed in Neufeld \& Melnick (1991), observations of these
transitions (and others that fall within the atmospheric windows) will
make it possible to use masers not only as a probe of dynamics, but
also of the physical conditions within the gas. Since these
transitions arise at different energies above the ground and have
different Einstein A values, the ratios of different masing
transitions will depend on the density and temperature. Furthermore,
by observing maser lines at different frequencies, it will be possible
to determine whether free-free absorption is affecting the
emission. For example, Herrnstein \etal (1996) proposed that the
asymmetry between the redshifted and blueshifted high-velocity
emission in NGC 4258 is not inherent, but is a consequence of the
geometry (see Figure 6): on the redshifted side, it is the
``underside'' (from our perspective) of the disk that is irradiated,
and so the warm, atomic zone lies behind the masing region, whereas on
the blueshifted side they are reversed, and the maser emission is
attenuated by free-free absorption within the weakly ionized atomic
zone. This hypothesis can be tested by observing the disk at
frequencies much higher than 22 GHz, where the optical depth to
free-free emission becomes negligibly small. By observing
density-sensitive line ratios, the actual densities in the masing zone
can be estimated, and compared with model predictions, such as the
NM95 model. David Neufeld and I are currently at work on improved
models of X-ray powered masers, to provide predictions of the line
strengths and ratios for future observations. 

\section*{Acknowledgments}
My research on masers is supported by the National Science Foundation
under grant AST 99-00871. This work was also supported by NASA through
HST grant AR-08747.02-A. I wish to thank James Moran and a second,
anonymous referee for their detailed and extremely helpful comments on
a draft of this review, and Lincoln Greenhill and Jim Braatz for
providing useful information. I am grateful to Brad Gibson and the ASA
meeting organizers for the invitation to present the talk on which
this article is based, and to Brad and Kristine and Joss and Sue for
their hospitality in Melbourne and Sydney, respectively. Thanks to Joe
for the company and all the driving in the 4WD monster and to Steve
for just about everything else, especially the sushi and beer. Above
all, I wish to thank Michelle Storey for her patience in dealing with
this extremely late manuscript.

\section*{References}

\reference Abramowicz, M.A., Chen, X., Kato, S., Lasota, J.-P., \&
Regev, O. 1995, ApJ, 438, L37

\reference Antonucci, R. 1993, ARAA, 31 473

\reference Baan, W.A., \& Haschick, A.D. 1996, ApJ, 473, 269 

\reference Balbus, S.A., \& Hawley, J.F. 1991, ApJ, 376, 214

\reference Braatz, J.A., Wilson, A.S., \& Henkel, C. 1997, ApJS, 110,
321 

\reference Braatz, J.A., Wilson, A.S., \& Henkel, C. 1996, ApJS, 106,
51 

\reference Bragg, A.E., Greenhill, L.J., Moran, J.M., \& Henkel,
C. 2000, ApJ, 535, 73

\reference Brock, D., Joy, M., Lester, D.F., Harvey, P.M., \& Ellis,
H.B., Jr. 1988, ApJ, 329, 208 

\reference Cecil, G., Bland-Hawthorn, J., Veilleux, S., \& Filippenko,
A.V. 2001, ApJ, 555, 338

\reference Cernicharo, J., Thum, C., Hein, H., John, D., Garcia, P., 
\& Mattioco, F. 1990, A\&A, 231, L15

\reference Cheung, A.C., Rank, D.M., Townes, C.H., Thornton, D.D.,
\& Welch, W.J. 1969, Nature, 221, 626

\reference Churchwell, E., Witzel, A., Huchtmeier, W., Pauliny-Toth,
I., Roland, J., \& Suben, W. 1977, A\&A 54, 969

\reference Claussen, M.J., \& Lo, K.-Y. 1986, ApJ, 308, 592

\reference Claussen, M.J., Heiligman, G.M., \& Lo,
K.-Y. 1984, Nature, 310, 298

\reference Claussen, M.J., Diamond, P.J., Braatz, J.A., Wilson, A.S.,
\& Henkel, C. 1998, ApJ, 500, L129 

\reference Collison, A.J., \& Watson, W.D. 1995, ApJ 452, L103

\reference Conway, J.E. 1999, in Highly Redshifted Radio Lines,
Ed. C.L. Carilli, S.J.E. Radford, K.M. Menten, \& G.I. Langston, (San
Francisco: Astronomical Society of the Pacific), 259

\reference Curran, S.J., Johansson, L.E.B., Rydbeck, G., \& Booth,
R.S. 1998, A\&A, 338, 863  

\reference de Jong, T. 1973, A\&A, 26, 297

\reference Desch, S.J., Wallin, B.K., \& Watson, W.D. 1998, ApJ,
496, 775

\reference Done, C., Madejski, G.M., \&  Smith, D.A. 1996, ApJ, 463,
L63 

\reference dos Santos, P.M., \& Lepine, J.R.D. 1979, Nature, 278, 34

\reference Duric, N., \& Seaquist, E.R. 1988, ApJ, 326, 574 

\reference Elitzur, M. 1992, Astronomical Masers, (Dordrecht: Kluwer),
262 

\reference Elmouttie, M., Haynes, R.F., Jones, K.L., Sadler, E.M.,
\& Ehle, M. 1998, MNRAS, 297, 1202 

\reference Fairall, A.P. 1986, MNRAS, 218, 453

\reference Falcke, H., Wilson, A.S., Henkel, C., Brunthaler, A., \&
Braatz, J.A. 2000a, ApJ, 530, L13

\reference Falcke, H., Henkel, C., Peck, A.B., Hagiwara, Y.,
Almudena Prieto, M., \& Gallimore, J.F. 2000b, A\&A, 358, L17

\reference Fiore, F., Pellegrini, S., Matt, G., Antonelli, L.A.,
Comastri, A.,  della Ceca, R., Giallongo, E., Mathur, S., Molendi, S.,
Siemiginowska, A., Trinchieri, G., \& Wilkes, B. 2001, ApJ, 556, 150

\reference Ford, H.C., Dahari, O., Jacoby, G.H., Crane, P.C., \&
Ciardullo, R. 1986, ApJ, 311, L7  

\reference Frank, J., King, A., \& Raine, D. 1992, Accretion Power in
Astrophysics (2nd ed.) (Cambridge: Cambridge University Press)

\reference Gallimore, J.F., Baum, S.A., \& O'Dea, C.P. 1997, Nature,
388, 852

\reference Gallimore, J.F., Baum, S.A., O'Dea, C.P., Brinks, E., \&
Pedlar, A. 1996, ApJ, 462, 740

\reference Gallimore, J.F., Henkel, C., Baum, S.A., Glass, I.S.,
Claussen, M.J., Prieto, M.A., \& Von Kap-herr, A. 2001, 556, 694

\reference Gammie, C.F., Narayan, R., \&  Blandford, R. 1999, ApJ,
516, 177

\reference Gardner, F.F., \& Whiteoak, J.B. 1982, MNRAS, 201, P13

\reference Genzel, R., \& Downes, D. 1979, A\&A 72, 234

\reference Greenhill, L.J. 2001a, in Cosmic Masers, ed. V. Migenes,
(San Francisco: Astronomical Society of the Pacific), in press.

\reference Greenhill, L.J. 2001b, in Proceedings of the 5th EVN 
Symposium, ed. J. Conway, A. Polatidis, R. Booth, Onsala Space
Observatory, Chalmers Technical University, Gothenburg, Sweden, in
press. 

\reference Greenhill, L.J., \& Gwinn, C.R. 1997, Ap\&SS, 248, 261

\reference Greenhill, L.J., Gwinn, C.R., Antonucci, R., \& Barvainis,
R. 1996, ApJ, 472, L21

\reference Greenhill, L.J., Moran, J.M., \& Herrnstein, J.R. 1997a,
ApJ, 481, L23

\reference Greenhill, L.J., Ellingsen, S.P., Norris, R.P., Gough,
R.G., Sinclair, M.W., Moran, J.M., \& Mushotzky, R. 1997b, ApJ, 474, L103

\reference Greenhill, L.J., Henkel, C., Becker, R., Wilson, T.L., \&
Wouterloot, J.G.A. 1995a, A\&A, 304, 21

\reference Greenhill, L.J., Jiang, D.R., Moran, J.M., Reid, M.J., Lo,
K.-Y., \& Claussen, M.J. 1995b, ApJ, 440, 619

\reference Greenhill, L.J., Ellingsen, S.P., Norris, R.P., Gough,
R.G., Sinclair, M.W., Moran, J.M., \& Mushotzky, R. 1997, ApJ, 474,
L103 

\reference Greenhill, L.J., Moran, J.M., Booth, R.S., Ellingsen, S.P.,
McCulloch, P.M., Jauncey, D.L., Norris, R.P., Reynolds, J.P.,
Tzioumis, A.K., \& Herrnstein, J.R. 2001, in Galaxies and their
Constituents at the Highest Angular Resolutions, ed. R. Schilizzi,
S. Vogel, F. Paresce, \& M. Elvis (San Francisco: Astronomical Society
of the Pacific), in press.

\reference Greenhill, L.J., Moran, J.M., \& Henkel, C. 2002, in
preparation. 

\reference Hagiwara, Y., Diamond, P.J., \& Miyoshi, M. 2002, A\&A,
383, 65 

\reference Hagiwara, Y., Kohno, K., Kawabe, R., \& Nakai, N. 1997,
PASJ, 49, 171 

\reference Haschick, A.D., \& Baan, W.A. 1985, Nature, 314, 144 

\reference Haschick, A.D., Baan, W.A., \& Peng, E.W. 1994, ApJ, 437, L35

\reference Heckman, T.M. 1980, A\&A, 87, 152 

\reference Henkel, C., G\"usten, R., Thum, C., \& Downes, D. 1984a, IAU
Circ. 3983 

\reference Henkel, C., G\"usten, R., Wilson, T.L., Biermann, P.,
Downes, D., \& Thum, C.,  1984b, A\&A, 141, L1

\reference Herrnstein, J.R. 1997, Ph.D. Thesis, Harvard University

\reference Herrnstein, J.R., Greenhill, L.J., \& Moran, J.M. 1996, ApJ,
468, L17

\reference Herrnstein, J.R., Moran, J.M., Greenhill, L.J., Blackman,
E.G., \& Diamond, P.J. 1998, ApJ, 508, 243

\reference Herrnstein, J.R., Greenhill, L.J., Moran, J.M., Diamond,
P.J., Inoue, M., Nakai, N., \& Miyoshi, M. 1998, ApJ, 497, L69

\reference Herrnstein, J.R., Moran, J.M., Greenhill, L.J., Diamond,
P.J., Miyoshi, M., Nakai, N., \& Inoue, M. 1997, ApJ, 475, L17

\reference Herrnstein, J.R., Moran, J.M., Greenhill, L.J., Diamond,
 P.J., Inoue, M., Nakai, N., Miyoshi, M., Henkel, C., \& Riess,
 A. 1999, Nature, 400, 539

\reference Ichimaru, S. 1977, ApJ, 214, 840

\reference Ishihara, Y., Nakai, N., Iyomoto, N., Makishima, K.,
Diamond, P., \& Hall, P. 2001, PASJ, 53, 215

\reference Iyomoto, N., Fukazawa, Y., Nakai, N., \& Ishihara, Y. 2001,
ApJ, 561, L69

\reference Iwasawa, K., Koyama, K., Awaki, H., Kunieda, H., Makishima,
K., Tsuru, T., Ohashi, T., \& Nakai, N. 1993, ApJ, 409, 155

\reference Kartje, J.F., K\"onigl, A., \& Elitzur, M. 1999, ApJ, 513, 180

\reference Koekemoer, A.M., Henkel, C., Greenhill, L.J., Dey, A.,
van Breugel, W., Codella, C., \& Antonucci, R. 1995, Nature, 378, 697

\reference Lasota, J.-P., Abramowicz, M.A., Chen, X., Krolik, J.,
Narayan, R., \& Yi, I. 1996, ApJ, 462, 142 

\reference Madejski, G.M., \. Zycki, P., Done, C., Valinia, A.,
Blanco, P., Rothschild, R., \& Turek, B. 2000, ApJ, 535, L87

\reference Makishima, K., Fujimoto, R., Ishisaki, Y., Kii, T.,
Loewenstein, M., Mushotzky, R., Serlemitsos, P., Sonobe, T., Tashiro,
M., \& Yaqoob, T. 1994, PASJ, 46, L77

\reference Maloney, P.R., Begelman, M.C., \& Pringle, J.E. 1996, ApJ,
472, 582

\reference Maloney, P.R., Hollenbach, D.J., \& Tielens, A.G.G.M. 1996,
ApJ 466, 561

\reference Maoz, E. 1995, ApJ, 447, L91

\reference Maoz, E., \& McKee, C. F. 1998, ApJ, 494, 218

\reference Marconi, A., Moorwood, A.F.M., Origlia, L., \& Oliva,
E. 1994, Messenger, 78, 20 

\reference Matt, G., Fiore, F., Perola, G.C., Piro, L., Fink, H.H.,
Grandi, P., Matsuoka, M., Oliva, E., \& Salvati, M. 1996, MNRAS, 281,
L69 

\reference Menten, K.M., Melnick, G.J., \& Phillips, T.G. 1990, ApJ,
350, L41

\reference Menten, K.M., Melnick, G.J., Phillips, T.G., \& Neufeld,
D.A. 1990, ApJ, 363, L27

\reference Miyoshi, M., Moran, J., Herrnstein, J., Greenhill, L.,
Nakai, N., Diamond, P., \& Inoue, M. 1995, Nature, 373, 127

\reference Moran, J.M., Greenhill, L.J., \& Herrnstein, J.R. 1999,
J. Astrophys. Astron., 20, 165 

\reference Moran, J., Greenhill, L., Herrnstein, J., Diamond, P.,
Miyoshi, M., Nakai, N., \& Inoue, M. 1995, in Quasars and AGN: High
Resolution Imaging, Proc. Nat. Acad. Sci., 92, 11427

\reference Nakai, N., Inoue, M., \& Miyoshi, M. 1993, Nature, 361, 45

\reference Nakai, N., Inoue, M., Miyazawa, K., \& Hall, P. 1995, PASJ
47, 771 

\reference Nakai, N., Sato, N., \& Yamauchi, A. 2002, PASJ, 54, in press.

\reference Narayan, R., \& Yi, I. 1994, ApJ, 428, L13

\reference Neufeld, D.A. 2000, ApJ, 542, L99

\reference Neufeld, D.A., \& Maloney, P.R. 1995, ApJ, 447, L19

\reference Neufeld, D.A., Maloney, P.R., \& Conger, S. 1994, ApJ, 436,
L127 

\reference Neufeld, D.A., \& Melnick, G.J. 1991, ApJ, 368, 215

\reference Newman, J.A., Ferrarese, L., Stetson, P.B., Maoz, E., Zepf,
 S.E., Davis, M., Freedman, W.L., \& Madore, B.F. 2001, ApJ, 553, 562  

\reference Phinney, E.S. 1983, Ph.D. Thesis, University of Cambridge

\reference Pringle, J.E. 1996, MNRAS, 281, 357

\reference Rees, M.J., Begelman, M.C., Blandford, R.D., \& Phinney,
E.S. 1982, Nature, 295, 17

\reference Reynolds, C.S., Nowak, M.A., \& Maloney, P.R. 2000, ApJ,
540, 143 

\reference Sawada-Satoh, S., Inoue, M., Shibata, K.M., Kameno, S.,
Migenes, V., Nakai, N., \& Diamond, P.J. 2000, PASJ, 52, 421

\reference Shakura, N.I., \& Sunyaev, R.A. 1973 A\&A 24, 337

\reference Stone, J.M., Hawley, J.F., Gammie, C.F., \& Balbus,
S.A. 1996, ApJ, 463, 656 

\reference Svensson, R. 1999, in Theory of Black Hole Accretion Discs,
ed. M.A. Abramowicz, G. Bjornsson, and J.E. Pringle,  (Cambridge:
Cambridge University Press), 289 

\reference Tielens, A.G.G.M., \& Hollenbach, D.J. 1985, ApJ, 291, 722

\reference Trotter, A.S., Greenhill, L.J., Moran, J.M., Reid,
M.J., Irwin, J.A., \& Lo, K.-Y. 1998, ApJ, 495, 470 

\reference Ulvestad J.S., Wrobel J.M., Roy A.L., Wilson, A.S.,
Falcke, H., \& Krichbaum, T.P. 1999, ApJ, 517, L81 

\reference Veilleux, S., \& Bland-Hawthorn, J., 1997 ApJ, 479, L105 

\reference Warwick, R.S., Koyama, K., Inoue, H., Takano, S., Awaki,
H., \& Hoshi, R. 1989, PASJ, 41, 709

\reference Waters, J.W., Kakar, R.K., Kuiper, T.B.H., Roscoe, H.K.,
Swanson, P.N., Rodriguez Kuiper, E.N., Kerr, A.R., Thaddeus, P.,
\& Gustincic, J.J. 1980, ApJ, 235, 57

\reference Weaver, K.A., Wilson, A.S., Henkel, C., \& Braatz,
J.A. 1999, ApJ, 520, 130

\reference Whiteoak, J.B. \& Gardner, F.F., 1986, MNRAS, 222, 513

\reference Xanthopoulos, E., \& Richards, A.M.S. 2001, MNRAS, 326, L37 

\end{document}